\documentclass[acmsmall, review=false, screen=true]{acmart}
\usepackage{booktabs} % For formal tables
\usepackage{makecell}
\usepackage{lipsum}
\usepackage[utf8]{inputenc}
\usepackage{array}
\usepackage{wrapfig}
\usepackage{multirow}
\usepackage{tabularx}
\usepackage{pifont}
\usepackage{lineno}
\usepackage{balance}
\usepackage{xcolor,pifont}
\usepackage{ragged2e}
\usepackage{hyphenat} % 添加断字支持
\usepackage{pifont} % 提供 \ding 命令

\usepackage{soul}
\newif\ifcleanversion
\cleanversiontrue
\ifcleanversion
  \DeclareRobustCommand{\rev}[1]{#1}
  
  \newcommand{\todo}[1]{}
  \newcommand{\revcolor}{}
\else
  \DeclareRobustCommand{\rev}[1]{\leavevmode{\color{blue}#1}}
  
  \newcommand{\todo}[1]{{\color{red}\textbf{TODO: #1}}}
  \newcommand{\revcolor}{\color{blue}}
\fi
\usepackage[linesnumbered, ruled,vlined]{algorithm2e}
\RequirePackage{pdflscape}
\usepackage{flushend}
\usepackage{multicol}
\usepackage{placeins}
\usepackage{booktabs}
\usepackage{indentfirst}
\usepackage{fancybox}
\usepackage[most]{tcolorbox}
\usepackage{fontawesome}
\usepackage{mathtools}
\usepackage{MnSymbol,wasysym}
\usepackage{rotating}
\usepackage{adjustbox}
\usepackage{colortbl}
\usepackage{color}
\usepackage{tcolorbox}
\usepackage{enumitem}
\newtcolorbox{mybox}[2][]
{colback = white, colframe = black, fonttitle = \bfseries,
    colbacktitle = gray, enhanced,
    attach boxed title to top left={yshift=-3mm, xshift=3mm},
    title=#2, #1}

\usepackage{framed}
\definecolor{Gray}{gray}{0.9}
\definecolor{shadecolor}{gray}{0.95}
\usepackage{tikz}
\usetikzlibrary{trees,positioning,shapes,shadows,arrows}
\tikzset{
  basic/.style  = {draw, text width=2cm, drop shadow, font=\sffamily, rectangle},
  root/.style   = {basic, rounded corners=2pt, thin, align=center, fill=white},
  level-2/.style = {basic, rounded corners=6pt, thin,align=center, fill=white, text width=3cm},
  level-3/.style = {basic, thin, align=center, fill=white, text width=1.8cm}
}
\usepackage[labelformat=simple]{subcaption} % 简化标签格式
\newcommand{\ours}[1]{\textsc{MAAD}}

\usepackage[normalem]{ulem}

\newtcolorbox[auto counter]{promptbox}[2][]{
  enhanced,
  width=\textwidth,
  colback=gray!10,
  colframe=blue!45!gray,
  boxrule=0.8pt,
  arc=2pt,
  left=8pt,
  right=8pt,
  top=6pt,
  bottom=6pt,
  colbacktitle=blue!45!gray,
  coltitle=white,
  fonttitle=\scriptsize\bfseries,
  title={Prompt \thetcbcounter: #2},  % This shows "1 Title" instead of "Title 1"
  label=#1,
  before upper={\scriptsize \ttfamily}
}

\hypersetup{colorlinks=true,linkcolor=blue,citecolor=blue,urlcolor=blue}
\ifcleanversion
  \hypersetup{linkcolor=black,citecolor=black,urlcolor=black}
\fi
\begin{document}
%\begin{sloppypar}

% \title{MAAD: Automate Software Architecture Design through Knowledge-Driven Multi-Agent Collaboration}
% \title{From Requirements to Blueprints: Knowledge-Driven Multi-Agent Collaboration for Software Architecture Design}
% \title{Knowledge-Driven Multi-Agent Collaboration for Software Architecture Design: Design, Implementation, and Empirical Study}
% \title{From Requirements to Blueprints: Knowledge-Augmented Multi-Agent Pipelines for End-to-End Architecture Design}
% \title{Towards Reliable Architecture Generation: A Knowledge-Augmented Multi-Agent Approach with Continuous Validation}
\title[Bridging Requirements and Architecture through Multi-Agent Orchestration]{Bridging Requirements and Architecture: Multi-Agent Orchestration with External Knowledge and Hierarchical Memory}

\author{Ruiyin Li}
% \authornotemark[1]
\email{ryli_cs@whu.edu.cn}
\orcid{0000-0001-8536-4935}
\affiliation{%
  \institution{School of Computer Science, Wuhan University}
  \city{Wuhan}
  \country{China}
}

\author{Yiran Zhang}
\email{yiran002@e.ntu.edu.sg}
\orcid{0000-0002-9366-6076}
\affiliation{
  \institution{Nanyang Technological University}
  \city{Singapore}
  \country{Singapore}
}

\author{Xiyu Zhou}
\email{xiyuzhou@whu.edu.cn}
\orcid{0009-0002-5946-0039}
\affiliation{
  \institution{School of Computer Science, Wuhan University}
  \city{Wuhan}
  \country{China}
}

\author{Yangxiao Cai}
\email{yangxiaocai@whu.edu.cn}
\orcid{0009-0007-7892-6611}
\affiliation{
  \institution{School of Computer Science, Wuhan University}
  \city{Wuhan}
  \country{China}
}

\author{Peng Liang}
\email{liangp@whu.edu.cn}
\orcid{0000-0002-2056-5346}
%\authornote{Corresponding author}
\affiliation{
  \institution{School of Computer Science, Wuhan University}
  \city{Wuhan}
  \country{China}
}

\author{Weisong Sun}
\email{weisong.sun@ntu.edu.sg}
\orcid{0000-0001-9236-8264}
\affiliation{
  \institution{Nanyang Technological University}
  \city{Singapore}
  \country{Singapore}
}

\author{Jifeng Xuan}
\email{jxuan@whu.edu.cn}
\orcid{0000-0002-2968-3496}
\affiliation{
  \institution{School of Computer Science, Wuhan University}
  \city{Wuhan}
  \country{China}
}

\author{Zhi Jin}
\email{zhijin@pku.edu.cn}
\orcid{0000-0003-1087-226X}
% \authornote{Corresponding author}
%\authornotemark[1]
\affiliation{
  \institution{School of Computer Science, Wuhan University}
  \city{Wuhan}
  \country{China}
}

\author{Yang Liu}
\email{yangliu@ntu.edu.sg}
\orcid{0000-0001-7300-9215}
\affiliation{
  \institution{Nanyang Technological University}
  \city{Singapore}
  \country{Singapore}
}

\thanks{This research is supported by the National Natural Science Foundation of China (NSFC) with Grant No. 92582203 and 62402348; National Research Foundation, Prime Minister's Office, Singapore under the Campus for Research Excellence and Technological Enterprise (CREATE) Programme; the National Research Foundation, Singapore, and DSO National Laboratories under the AI Singapore Programme (AISG Award No: AISG2-GC-2023-008). The authors would also like to thank the architects who participated in the interviews in this study.}

\acmJournal{TOSEM}
\acmVolume{0}
\acmNumber{0}
\acmArticle{0}
\acmMonth{0}

\renewcommand{\shortauthors}{Li et al.}

\begin{abstract}
Software architecture design is a critical yet inherently complex and knowledge-intensive phase that requires balancing competing quality attributes and adapting to evolving requirements. Traditionally, this process has been time-consuming, labor-intensive, and heavily reliant on architects, often resulting in limited exploration of alternative architectural decompositions and styles, especially under the pressures of agile development. While Large Language Model (LLM)-based agents have shown promising performance across various software engineering tasks, \rev{coordinating requirements analysis, multi-view modeling, and design documentation still requires explicit attention to consistency and decision rationale.} In addition, single-agent approaches often yield architectural outputs with inconsistent cross-view designs or incomplete requirements coverage, and existing multi-agent systems lack reasoning grounded in domain-specific architectural constraints and quality requirements. To address these challenges, we propose MAAD (Multi-Agent Architecture Design), a knowledge-driven framework that orchestrates four specialized agents (i.e., Analyst, Modeler, Designer, and Evaluator) to autonomously \rev{transform requirements specifications into multi-view architectural models, documentation, and preliminary evaluation reports}. MAAD incorporates Retrieval-Augmented Generation (RAG) to inject recognized architectural standards and patterns into the workflow and leverages a hierarchical memory mechanism that captures design history for iterative refinement. We evaluate MAAD through comparative experiments against MetaGPT, using quantitative architecture-level metrics across 10 case studies, \rev{a comparison of configurations with and without external knowledge, a comparison of four foundational LLMs, and interviews with six practitioners}. \rev{Results show that MAAD produces more detailed architectural descriptions together with dedicated evaluation reports than the selected MetaGPT outputs, with differences in the reported structural indicators; the effects of knowledge infusion vary across metrics and cases.} \rev{Furthermore, the quality of the generated artifacts depends on the underlying LLM's reasoning capacity, with GPT-5.2 and Qwen3.5 producing more detailed artifacts than the other compared models. Practitioners regard the outputs as useful initial design drafts and emphasize human validation of requirements, trade-offs, and approval.} The replication package of MAAD is available at~\cite{onlinepackage_TOSEM} to support reproducibility and future extensions.
\end{abstract}
% [---- For Submission Only -----]
% 

% [CCS的分类] https://dl.acm.org/ccs

\maketitle

\section{Introduction}\label{sec:Introduction}
Software architecture design lies at the heart of any successful software project, defining high-level system structures, allocating responsibilities, and prescribing interactions that satisfy both functional and quality attributes~\cite{Bass2021SAP}. \rev{In practice, architects interpret stakeholder concerns, explore alternatives, and justify choices under technical and organizational constraints. Architectural models communicate those choices, and decision rationale records why a given alternative fits the project; both depend on sustained communication with stakeholders and development teams, not on blueprints alone~\cite{Kruchten2008ReallyDo,Kruchten2009DecisionView}.}

\rev{Large Language Models (LLMs) are increasingly being investigated for architecture modeling, architectural knowledge retrieval, and design-decision support~\cite{Schmid2025LLMSA,GenAI4SA2026,dhar2024can}. Recent approaches include requirements-to-architecture mapping~\cite{Helmi2025ARLO}, Attribute-Driven Design with human collaboration~\cite{Cervantes2026ADD}, and collaborative LLM agents for C4 model generation~\cite{Szczepanik2026C4}. These efforts target specific architecture activities. Coordinating requirements analysis, multi-view architecture modeling, documentation, and iterative review within an explicit architecture-oriented workflow remains an open problem.}

\rev{Here, an LLM-based agent refers to a role-specific component that leverages an LLM, task instructions, and contextual artifacts to perform a designated activity. A multi-agent system (MAS) is an orchestration of such generative agents and their artifact exchanges, rather than a system based on classical multi-agent coordination algorithms. By assigning agents to specialized roles and enabling them to reason independently and communicate iteratively, an MAS emulates the collaborative process of a human development team toward a shared goal. General-purpose frameworks such as MetaGPT organize software development through roles and standard operating procedures~\cite{hong2023metagpt,He2025mas}. Their default targets are code-centric deliverables; multi-view architecture modeling, decision rationale, and architecture evaluation are not primary objectives. Therefore, an architecture-oriented workflow is essential to support its own roles, intermediate artifacts, and review loop.}

\rev{Orchestrating LLM agents for architecture design raises three challenges that follow from how generative models produce and reuse text. First, requirements may be omitted, reinterpreted, or assigned inconsistent identifiers as they pass between agents and artifacts. Second, separately generated architecture views and documents may disagree on responsibilities, interfaces, and constraints even when they address the same system. Third, retrieved architectural guidance and accumulated design history are not automatically grounded in the current requirements; agents may reproduce familiar architectural patterns or detailed diagrams without establishing project-appropriate decisions. These challenges call for explicit intermediate architectural artifacts, iterative evaluation feedback, and requirement-grounded architectural knowledge organization.}

\rev{To this end, we propose \textbf{Multi-Agent Architecture Design (MAAD)}, a knowledge-driven framework for generating and reviewing early-stage architectural artifacts. MAAD orchestrates four roles: the \textit{Analyst} structures requirements, the \textit{Modeler} constructs multi-view models, the \textit{Designer} synthesizes architecture documentation, and the \textit{Evaluator} supplies feedback and preliminary evaluation reports. Retrieval-Augmented Generation (RAG)~\cite{Lewis2020RAG} supplies architectural standards and guidance, while working, episodic, and semantic memory organize intermediate design artifacts, revision history, and reusable architectural design rationale across iterations~\cite{Zhang2025MMS}. The contribution is this integrated architecture-oriented workflow, not RAG or role specialization alone.}

\rev{We evaluate MAAD against MetaGPT, examine external-knowledge and model configurations, and gather practitioner feedback on the generated drafts and review reports, showing that MAAD produces structured early-stage architectural artifacts that practitioners view as useful design and review aids. Consistent with Ivers and Ozkaya~\cite{Ivers2025AutomationGap}, we position MAAD for automated artifact generation rather than replacing stakeholder-driven validation and approval. This manuscript extends our preliminary vision paper~\cite{Zhang_MAAD2025} with an implemented framework, comparative experiments, and practitioner interviews. Its contributions are threefold:}
\begin{itemize}
\item \rev{\textbf{Architecture-oriented agent workflow.} Four agents cover requirements analysis, multi-view architecture modeling, documentation synthesis, and iterative review, with explicit responsibilities and artifact exchanges.}
\item \rev{\textbf{Knowledge and memory organization.} External architectural knowledge is combined with three memory layers that organize artifacts, feedback, and design history within the workflow.}
\item \rev{\textbf{Empirical evaluation.} We compare MAAD's generated artifacts and structural metrics against MetaGPT as well as across knowledge-infusion and foundational-model configurations, and report practitioner feedback on how these outputs fit into human-led architecture work.}
\end{itemize}

\textbf{Organization}: Section~\ref{sec:RelatedWork} introduces related studies of our work. Section~\ref{sec:MAADFramework} elaborates on the design of the MAAD framework, and Section~\ref{sec:StudyDesign} defines the research questions. Section~\ref{sec:Results} presents the results and our findings, and Section~\ref{sec:Discussions_and_Implications} discusses the results of this study. Section~\ref{sec:Threats} examines the threats to the validity of this study. Section~\ref{sec:Conclusion} summarizes this study and outlines the future work.

\section{Related Work}\label{sec:RelatedWork}
\subsection{Software Architecture Design}
Software architecture design has evolved significantly over the past few decades, transitioning from experience-based heuristics to systematic, model-driven engineering practices~\cite{Bass2021SAP}. As systems grew in complexity, researchers introduced Architecture Description Languages (ADLs) to bring formality to architectural modeling and analysis. Prominent ADLs such as AADL~\cite{Feiler2006AADL} allow researchers and practitioners to define system components, connectors, and configurations systematically~\cite{Medvidovic2000ADL}. Model-Driven Architecture (MDA) further emphasized the transformation of abstract architectural models into platform-specific implementations through model transformations, bridging the gap between abstract design and executable systems~\cite{Mellor2004MDAD}. \rev{Building on model-driven engineering (MDE), Perovich \textit{et al.}~\cite{Perovich2009MDE} automate architectural design transformations from explicit feature and structural models. Thus, ADLs and MDE already provide substantial automation once formal source models and transformation rules are defined.} %; MAAD addresses a complementary setting by generating and iteratively refining early-stage architectural artifacts directly from natural-language specifications.}

In parallel, architectural decision modeling and quality attribute-driven evaluation approaches like Architecture Tradeoff Analysis Method (ATAM)~\cite{Kazman2000ATAM} have provided systematic ways to align architectural choices with business goals and non-functional requirements~\cite{Bass2021SAP}. \rev{Specifically, Kruchten's ``4+1'' view model organizes complementary structural, behavioral, development, and deployment perspectives~\cite{Kruchten1995avm}, while the decision view captures architectural choices, alternatives, constraints, and their underlying rationale~\cite{Kruchten2009DecisionView}. Kruchten~\cite{Kruchten2008ReallyDo} further emphasizes that architects do more than produce blueprints, dedicating substantial effort to communicating with stakeholders and development teams. Following this perspective, MAAD pairs ``4+1'' multi-view models with explicit rationale documentation, and treats its ATAM-inspired evaluation report as preliminary analysis material rather than a substitute for stakeholder-driven ATAM workshops~\cite{Kazman2000ATAM}.}

\subsection{Large Language Models for Software Architecture Design}
\rev{Large Language Models (LLMs) have received significant attention across software development~\cite{li2025ChatGPT}, and recent reviews~\cite{Schmid2025LLMSA,GenAI4SA2026} show growing adoption across architectural tasks such as architectural pattern detection and decision support. At the same time, empirical evidence indicates that off-the-shelf LLMs remain prone to hallucinations and incomplete architectural knowledge, calling for verified knowledge sources and structured evaluation frameworks~\cite{GenAI4SA2026,Soliman2025Knowledge}.}

\rev{Most existing approaches target individual architecture activities along the design pipeline. For requirements and visual modeling, prior studies examine LLM-assisted requirements drafting~\cite{Pasquale2025Requirements}, early requirements-to-architecture mapping~\cite{Eisenreich_2024}, and UML diagram generation and assessment~\cite{Wang2024UML,Nguyen2026Class}. While conversational LLMs (e.g., ChatGPT) paired with notations such as PlantUML and Mermaid simplify standalone diagram rendering, they neither enforce cross-view consistency nor tie models to design rationale. For architectural decision-making,} Dhar \textit{et al}.~\cite{dhar2024can} found that zero-shot LLM generation of architecture decision records still falls short of human reasoning. \rev{More structured methods address this gap by coupling LLMs with explicit decision frameworks: ARLO~\cite{Helmi2025ARLO} maps requirements to a configurable choice matrix solved via integer linear programming, and Cervantes \textit{et al.}~\cite{Cervantes2026ADD} embed LLMs into Attribute-Driven Design (ADD) with human guidance. Recent preprints also explore LLM support specifically for quality attributes: Capilla \textit{et al.}~\cite{Capilla2026QAEval} evaluate LLM-assisted analysis of ATAM quality-attribute scenarios (identifying risks, sensitivity points, and trade-offs), while QUARE~\cite{Cheng2026QUARE} uses quality-specialized LLM agents with dialectical negotiation to balance competing quality attributes during requirements engineering. Yet these methods focus on specific artifacts rather than coordinating requirements, multi-view models, documentation, and review end to end.}

\rev{LLM-based multi-agent systems (MAS) offer a natural mechanism for coordinating such multi-step SE activities~\cite{He2025mas,Cai2026MAS}. However, general-purpose frameworks such as MetaGPT~\cite{hong2023metagpt} target code-centric deliverables, and role-simulation libraries such as TinyTroupe~\cite{TinyTroupe} lack architecture-specific artifact contracts. In concurrent work, Szczepanik and Chudziak~\cite{Szczepanik2026C4} apply collaborative LLM agents to generate and check C4 views, focusing primarily on structural diagrams. Extending our preliminary vision~\cite{Zhang_MAAD2025}, MAAD instead coordinates requirements structuring, ``4+1'' multi-view architecture modeling, rationale documentation, and iterative evaluation through external architectural knowledge and hierarchical memory, aligning with recent calls to ground GenAI support in concrete architecting workflows~\cite{Ivers2025AutomationGap}.}

\subsection{Conclusive Summary}
Table~\ref{T:Comparison} summarizes the representative studies, their scope and gaps, and comparisons with our work. The table is organized into four columns: \textit{Theme}, \textit{Previous Studies}, \rev{\textit{Scope and Gap}}, and \textit{Comparison}.

Overall, prior studies on architecture design and LLM-based SE can be broadly grouped into three themes. First, traditional software architecture research focuses on modeling methodologies, \rev{MDE transformations,} and evaluation frameworks, \rev{which automate structured model transformations but require upfront formalization or substantial manual effort when starting from unstructured natural-language requirements.} Second, recent LLM-based approaches have demonstrated promising capabilities in specific architectural tasks \rev{(e.g., requirement drafting, UML diagramming with PlantUML/Mermaid, and ADR generation)}, yet lack an integrated workflow that connects multi-view architecture modeling, documentation, and iterative review. Third, although multi-agent systems \rev{and role-definition libraries} support task decomposition and collaboration, existing MAS frameworks are either code-centric \rev{or focused on specific structural views (e.g., C4 models), without combining ``4+1'' modeling, rationale documentation, external architectural knowledge, and iterative quality-attribute review.}

\rev{To bridge this workflow gap,} we propose MAAD, a knowledge-driven multi-agent framework for generating and reviewing early-stage architectural artifacts. By orchestrating specialized agents, integrating retrieval-augmented domain knowledge, and enabling iterative evaluation with hierarchical memory, MAAD connects requirements analysis, multi-view architecture modeling, documentation, and review within an explicit architecture-oriented workflow.

\begin{table}[hptb]
\centering
\small
\caption{Comparison of Previous Studies and Our Work}\label{T:Comparison}
\begin{tabularx}{\linewidth}{>{\RaggedRight\arraybackslash}p{2.2cm} *{3}{>{\RaggedRight\arraybackslash}X}}
\toprule
\textbf{Theme} & \textbf{Previous Studies} & \textbf{\rev{Scope and Gap}} & \textbf{Comparison} \\
\midrule
\textbf{Software Architecture Design} 
& Systematic design, modeling, and evaluation, including ADLs such as AADL~\cite{Feiler2006AADL,Medvidovic2000ADL}, model-driven approaches~\cite{Mellor2004MDAD,Perovich2009MDE}, \rev{``4+1'' and decision views~\cite{Kruchten1995avm,Kruchten2009DecisionView},} and ATAM~\cite{Kazman2000ATAM}. 
& \rev{ADLs and MDE automate transformations over formal models, but require upfront formalization and intensive expert effort to bridge natural-language requirements, rationale, and evaluation.} 
& \rev{Unlike MDE solutions that transform explicit source models, MAAD automates early-stage artifact generation and review directly from natural-language requirements while preserving multi-view and decision principles.} \\
\addlinespace[4pt]
\textbf{LLMs for Software Engineering and Architecture}
& LLMs for SE~\cite{li2025ChatGPT}, \rev{requirements drafting~\cite{Pasquale2025Requirements}, UML modeling~\cite{Wang2024UML,Nguyen2026Class} with notations (PlantUML, Mermaid), knowledge/ADR tasks~\cite{Soliman2025Knowledge,dhar2024can}, decision-guided design (ARLO~\cite{Helmi2025ARLO}, ADD~\cite{Cervantes2026ADD}), and quality-attribute evaluation and negotiation~\cite{Capilla2026QAEval,Cheng2026QUARE}.} 
& Mostly task-specific or single-agent; \rev{conversational LLMs (e.g., ChatGPT) with PlantUML or Mermaid render standalone diagrams without ensuring cross-view consistency, rationale, or structured review.} 
& \rev{Beyond standalone diagramming with conversational LLMs and notations (PlantUML/Mermaid) or single-task support,} MAAD provides a unified pipeline from requirements to multi-view architecture modeling, documentation, and iterative evaluation. \\
\addlinespace[4pt]
\textbf{Multi-Agent Systems for Software Engineering} 
& MAS frameworks (e.g., MetaGPT~\cite{hong2023metagpt}, \rev{SE agent workflows~\cite{He2025mas,Cai2026MAS}), role-simulation libraries (e.g., TinyTroupe~\cite{TinyTroupe}), and collaborative C4 agents~\cite{Szczepanik2026C4}.} 
& General MAS are code-centric, \rev{persona libraries lack architectural contracts, and C4 agents focus on structural views without full ``4+1'' documentation, knowledge grounding, and trade-off review.} 
& \rev{Rather than generic role libraries (e.g., TinyTroupe) or code-centric pipelines (e.g., MetaGPT),} MAAD couples architecture-specific roles with RAG-based domain knowledge, hierarchical memory, and an \textit{Evaluator} agent for iterative review. \\
\bottomrule
\end{tabularx}
\end{table}

\section{MAAD Framework}\label{sec:MAADFramework}
In this section, we present the MAAD framework in three parts. First, we provide an overview of the MAAD framework in Section~\ref{sec:Overview}. Second, we describe the design details of the constituent agents of MAAD in Section~\ref{sec:AgentDesign}. Third, we introduce the hierarchical memory mechanism for the MAAD framework in Section~\ref{sec:MemoryMechanism}. 

\subsection{Overview}\label{sec:Overview}
\begin{figure}[t]
    \centering
    \includegraphics[width=\linewidth]{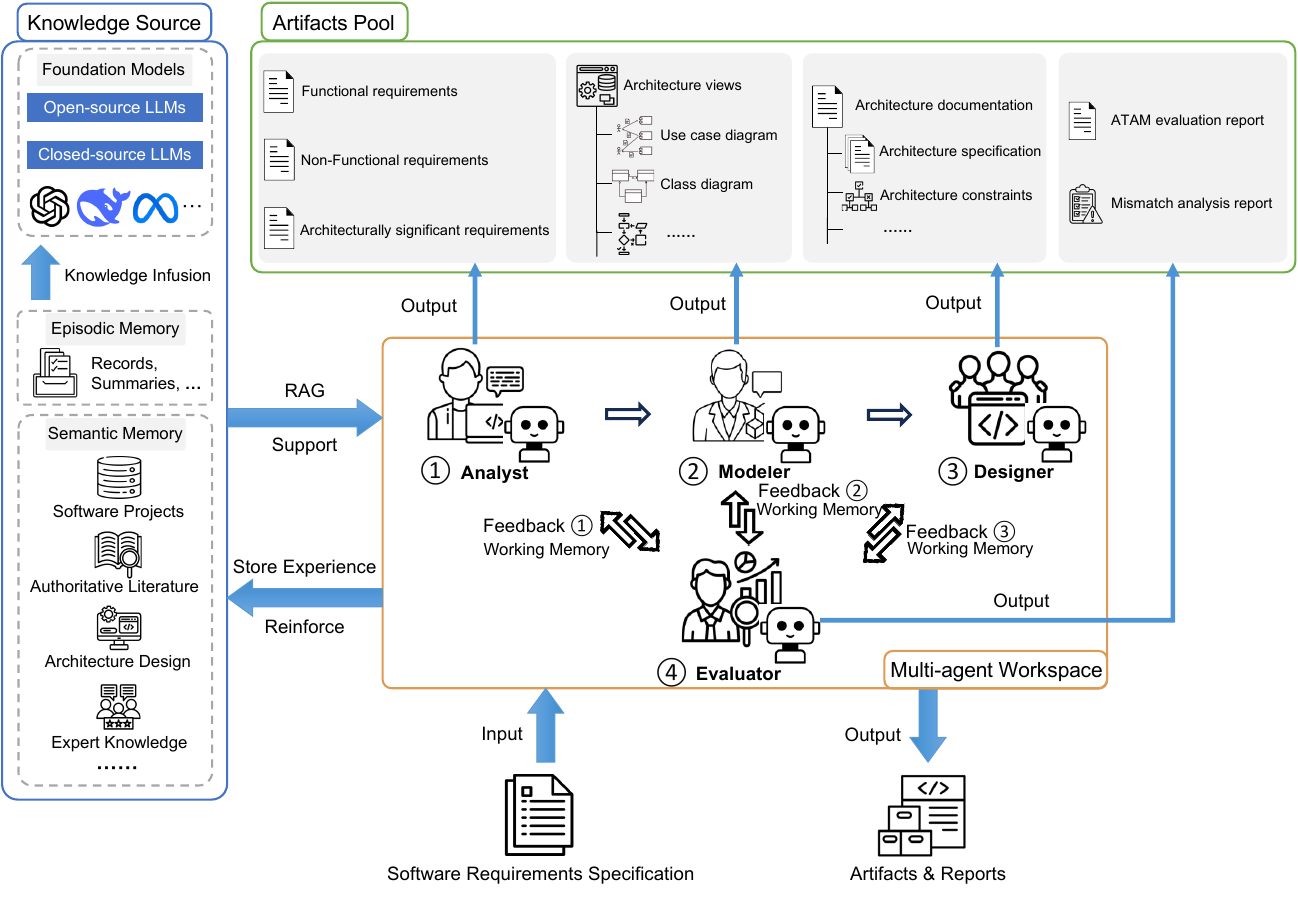}
    \caption{Overview of the MAAD Framework}\label{F:Overview}
\end{figure}

The MAAD framework implements a knowledge-driven, iterative, multi-agent pipeline that autonomously transforms a Software Requirements Specification (SRS) into \rev{a candidate architecture description} (see Figure~\ref{F:Overview}). MAAD comprises four specialized agents---\textit{Analyst}, \textit{Modeler}, \textit{Designer}, and \textit{Evaluator}---each equipped with perception, reasoning, memory, and action capabilities. These agents interact through a shared artifact pool, draw on external knowledge infusion, \rev{and remain model-agnostic so that the pipeline can run on either proprietary or open-sourced foundation LLMs (evaluated in Section~\ref{sec:LLMSelection}). Although Section~\ref{sec:AgentDesign} details the agents in workflow order, we highlight the \textit{Evaluator} upfront because it acts as a cross-stage quality gate throughout the pipeline:} it iteratively assesses intermediate requirements, models, and documentation for requirement alignment and internal consistency, and finally produces system-level preliminary evaluation reports.

\rev{Rather than an arbitrary split, the four roles mirror core activities in established architecture workflows~\cite{ISO_IEC_IEEE_42020_2019,Bass2021SAP,Zhang_MAAD2025}, mapping each major intermediate deliverable to a dedicated agent (see Table~\ref{T:agent_roles}). Specifically, the \textit{Analyst} structures raw requirements; the \textit{Modeler} performs architectural abstraction, selecting patterns and tactics to construct multi-view UML models; and the \textit{Designer} translates those high-level views into developer- and operations-facing documentation, including interface contracts, data schemas, deployment specifications, and decision rationale. Separating the \textit{Modeler} from the \textit{Designer}---instead of tasking one agent with generating diagrams, full technical documentation, and self-checks simultaneously---prevents loss of focus in long contexts across ten diagram categories and multi-section specifications, enables targeted patch-based revisions on models versus text, and avoids self-validation bias by delegating quality checks to an independent \textit{Evaluator}.}

\begin{table}[htbp]
\centering\small
\caption{\rev{Inputs, core responsibilities, and main outputs of the four MAAD agents}}\label{T:agent_roles}
\begin{tabularx}{\linewidth}{@{}l *{3}{>{\RaggedRight\arraybackslash}X}@{}}
\toprule
\rev{\textbf{Agent}} & \rev{\textbf{Primary Inputs}} & \rev{\textbf{Core Responsibility}} & \rev{\textbf{Main Outputs}}\\
\midrule
\rev{\textit{Analyst}} & \rev{Raw SRS, semantic memory} & \rev{Extract requirements and isolate architectural drivers} & \rev{Structured FR, NFR, and ASR artifacts}\\
\addlinespace[3pt]
\rev{\textit{Modeler}} & \rev{Validated FR/NFR/ASRs, RAG knowledge, memory} & \rev{Select patterns/tactics and construct ``4+1'' views} & \rev{Multi-view PlantUML diagrams and QA trade-off summary}\\
\addlinespace[3pt]
\rev{\textit{Designer}} & \rev{Validated FR/NFR/ASRs, UML views, RAG knowledge, memory} & \rev{Synthesize technical/deployment specs, traceability, and rationale} & \rev{Draft architecture document (APIs, schemas, deployment, rationale)}\\
\addlinespace[3pt]
\rev{\textit{Evaluator}} & \rev{Stage drafts (in-loop); SRS and all artifacts (final)} & \rev{Validate stage outputs for refinement and run system-level analysis} & \rev{Stage feedback JSONs, mismatch report, and ATAM-inspired report}\\
\bottomrule
\end{tabularx}
\end{table}

As shown in Figure~\ref{F:Overview}, the workflow begins when the system receives the SRS as input. (1) The \textbf{Analyst agent} analyzes the SRS to identify and extract key requirement elements, and decomposes the specification into structured requirements artifacts. These artifacts are subsequently assessed by the \textit{Evaluator} agent to ensure the completeness of requirements coverage, correctness of requirements interpretation, and consistency among extracted requirements and domain constraints. (2) The \textbf{Modeler agent} transforms analyzed requirements into multi-view architectural models based on ``4+1'' architecture view models proposed by Kruchten \cite{Kruchten1995avm}, which is a widely used architecture description. The resulting artifacts are again evaluated by the \textit{Evaluator} agent to verify their quality and alignment with the requirements artifacts. (3) The \textbf{Designer agent} perceives the validated artifacts generated by both the \textit{Analyst} and \textit{Modeler} agents. Using these inputs, it synthesizes the final architecture documentation. Likewise, the \textit{Evaluator} agent performs quality assessment on the produced artifacts before they are finalized. (4) Throughout the process, the \textbf{Evaluator agent} continuously assesses artifacts at each stage and performs a final system-level evaluation, producing two key outputs: \rev{an ATAM-inspired preliminary evaluation report} and a mismatch analysis report. These reports highlight architectural trade-offs and identify potential inconsistencies or deviations between the generated architecture artifacts and the original requirements.

The process is an iterative and feedback-driven workflow. At each stage, artifacts are evaluated before being accepted, and \rev{feedback is used to request revisions. Completion of this loop is not evidence of satisfaction of runtime quality attributes.} MAAD incorporates a hierarchical memory mechanism~\cite{Zhang2025MMS}. Each agent maintains a \textit{working memory} that stores the current architecture design context, intermediate artifacts, and evaluation feedback required for ongoing reasoning and decision-making during iterative design activities. In addition, each agent maintains an \textit{episodic memory} that records architectural design experiences generated during agent interactions. At the framework level, a shared \textit{semantic memory} stores generalized architectural knowledge distilled from past knowledge sources (e.g., software projects, architecture design references, authoritative literature, and expert knowledge), and all agents can access and retrieve the relevant architectural knowledge from the shared \textit{semantic memory}. \rev{This memory mechanism organizes knowledge for reuse and makes earlier feedback available during subsequent refinement stages.} As a result, the collaborative multi-agent process enables the MAAD framework to synthesize \rev{a candidate architecture description with model-generated review feedback} from the initial SRS.

\rev{We use \textit{autonomous execution} to describe the generation mode: agents perform their assigned tasks and exchange refinement feedback without requiring a human to author each intermediate artifact. This execution mode remains compatible with human oversight. Stakeholders validate the requirements and their priorities, and architects review the generated alternatives, rationale, and risks before accepting a design. Section~\ref{sec:Implications} further explains where these responsibilities enter the broader architecting workflow.}

\subsection{Agent Design}\label{sec:AgentDesign}

\subsubsection{Analyst Agent}\label{sec:AnalystAgent}
The \textit{Analyst} agent is responsible for interpreting the Software Requirements Specification (SRS) and producing structured requirements artifacts. Specifically, it produces three types of requirements artifacts: \textit{Functional Requirements (FRs)}, \textit{Non-Functional Requirements (NFRs)}, and \textit{Architecturally Significant Requirements (ASRs)}. These artifacts are stored in the shared artifact pool and subsequently consumed by downstream agents in the architecture design pipeline.

The \textit{Analyst} agent performs requirements analysis through a structured reasoning process guided by prompts and supported by external knowledge (see Section~\ref{sec:AgentSetting}). It first parses the SRS to extract requirement statements expressed in natural language, which are then classified into FRs and NFRs. FRs describe the system's functional behavior, while NFRs capture QAs and constraints, such as performance, reliability, security, and scalability. \rev{Importantly, architectural significance is an orthogonal property rather than a mutually exclusive third category alongside FRs and NFRs~\cite{Bass2021SAP}: both FRs and NFRs can qualify as ASRs whenever they shape core module decomposition, impose cross-cutting constraints, or trigger consequential QA trade-offs (e.g., a strict latency target affecting caching and deployment, unlike a local UI label preference). Building on the extracted FRs and NFRs, the agent isolates this architecturally significant subset into a dedicated ASR artifact so that downstream agents explicitly prioritize key architectural drivers.}

\textbf{Prompt Design}. The \textit{Analyst} agent uses two structured prompts to extract and refine requirements from the SRS\rev{, each executed as a single unified prompt rather than separate per-rule sub-prompts.} \texttt{Prompt~1} encodes domain knowledge about requirements engineering and architectural analysis while enforcing a deterministic output structure. \rev{Beyond listing high-level ASR indicators, the full prompt grounds ASR identification by combining concrete requirement templates, injected reference knowledge, and semantic memory with mandatory per-ASR justification fields---requiring the agent to explicitly record the \textit{architectural impact} (e.g., on decomposition, communication patterns, or technology choices), \textit{affected QAs}, \textit{architectural constraints}, \textit{rationale}, and \textit{dependencies/conflicts}.} \texttt{Prompt~2} supports iterative architecture refinements based on feedback from the \textit{Evaluator} agent (see Section~\ref{sec:EvaluatorAgent}). To ensure consistency and machine-readability of generated artifacts, the prompts explicitly restrict LLMs to generate outputs that strictly conform to a predefined Markdown format, without additional explanations or reasoning traces. This constraint reduces output variability and ensures that generated artifacts can be consistently parsed and consumed by downstream agents. \rev{Note that the prompt boxes throughout Section~\ref{sec:AgentDesign} present condensed summaries of core task instructions and output schemas; all verbatim system prompts (including the \textit{Evaluator}'s stage-level review prompts) are provided in our replication package~\cite{onlinepackage_TOSEM}.}

\textbf{Iterative Refinement and Memory Integration}: The \textit{Evaluator} agent assesses the artifacts generated by the \textit{Analyst} agent \rev{via a dedicated requirement-review prompt enforcing a multi-point checklist (covering ASR relevance, FR atomicity, NFR measurability, ambiguity, testability, and inter-requirement conflicts) and returns structured JSON feedback (\texttt{issues\_found}) specifying affected \texttt{requirement\_id}s, severity levels, and concrete revision suggestions.} The \textit{Analyst} agent selectively refines the identified issues in subsequent iterations \rev{via \texttt{Prompt~2}: for flagged conflicts, compatible numeric targets are reconciled into a unified entry with the original requirement IDs recorded under \texttt{Dependencies/Conflicts}, whereas irreconcilable items are marked \texttt{DEPRECATED} alongside a reconciled replacement while all unflagged requirements remain verbatim.} This feedback-driven refinement process improves the clarity, completeness, and architectural relevance of requirements artifacts, ensuring their suitability for downstream architectural modeling and architectural design.

\begin{promptbox}[box:prompt1]{Simplified Prompt of the \textit{Analyst} Agent for Requirements Extraction}
\textbf{Role}: Requirements analysis expert.\\
\textbf{Input}: Requirements Document (SRS); Domain Knowledge Reference; Past Architectural Lessons (Semantic Memory).\\
\textbf{Task}: 1. Extract Functional Requirements (FR); 2. Extract Non-Functional Requirements (NFR); 3. Identify Architecturally Significant Requirements (ASR).\\
\textbf{Classification Rules}: 1. FRs describe system behaviors and operations; 2. NFRs describe quality attributes and operational constraints; 3. ASRs are requirements that significantly influence architectural decisions.\\
\textbf{Indicators of ASRs include}: strong architectural constraints; cross-cutting concerns; quality attribute trade-offs; high-risk or high-value requirements.\\
\textbf{Output}: Generate three structured artifacts in Markdown format.
\end{promptbox}

\begin{promptbox}[box:prompt2]{Simplified Prompt of the \textit{Analyst} Agent for Requirements Refinement}
\textbf{Role}: Requirements refinement expert.\\
\textbf{Input}: Requirements Document (SRS); Current FR/NFR/ASR Artifacts; Evaluator Feedback (JSON); Domain Knowledge Reference; Past Architectural Lessons (Semantic Memory).\\
\textbf{Task}: Apply evaluator feedback to refine Functional Requirements (FR), Non-Functional Requirements (NFR), and Architecturally Significant Requirements (ASR).\\
\textbf{Refinement Rules}: 1. Process each issue in Evaluator feedback sequentially; 2. Modify only the requirement identified by \texttt{requirement\_id}; 3. Preserve JSON schema structure, IDs, formatting, and traceability; 4. Do not invent or infer missing information; 5. Split, merge, deprecate, or reword requirements only when explicitly instructed; 6. Preserve all unrelated content verbatim.\\
\textbf{Special Handling}: 1. Add lineage notes when requirements are decomposed; 2. Resolve conflicts only as instructed; 3. Include owner/next-action tags if provided in feedback.\\
\textbf{Output}: Generate updated FR, NFR, and ASR artifacts in the original Markdown format.
\end{promptbox}

\subsubsection{Modeler Agent}\label{sec:ModelerAgent}
The \textit{Modeler} agent transforms analyzed requirements into a structured, multi-view architectural representation that bridges requirements and system design. Within the MAAD pipeline, the \textit{Modeler} agent utilizes architectural knowledge through the RAG technique~\cite{Lewis2020RAG} \rev{(see Section~\ref{sec:AgentSetting})} to construct ``4+1'' view models~\cite{Kruchten1995avm}, including \textit{Scenario View}, \textit{Logical View}, \textit{Process View}, \textit{Development View}, and \textit{Physical View}. These view models are generated as \textit{PlantUML} diagrams, enabling both human interpretability and machine verifiability. \rev{Generating all five perspectives establishes full multi-view coverage and enables cross-view consistency checking across structural, behavioral, development, and deployment boundaries, while architects remain free to inspect or present only the subset of views relevant to specific stakeholders. Rather than replacing ``4+1'' with a standalone decision view, MAAD treats structural/behavioral views (\textit{what} the architecture is) and decision rationale (\textit{why} options are chosen)~\cite{Kruchten2009DecisionView} as complementary across the \textit{Modeler} and \textit{Designer}.}

Guided by structured prompts and supported by semantic memory, the \textit{Modeler} agent performs architecture modeling through a systematic reasoning process. First, \rev{without hardcoding domain-specific priorities,} it analyzes the project's NFRs and ASRs to identify key QAs (e.g., performance, scalability) and potential trade-offs among QAs, selects appropriate architectural patterns (e.g., layered, microservices), and derives architectural patterns and tactics that address specific quality concerns. Subsequently, the \textit{Modeler} agent derives architectural elements (e.g., domain entities, components) and organizes them into the ``4+1'' views. Each view adheres to strict rules (e.g., naming, traceability, and cross-view consistency), ensuring naming consistency and traceability to requirements. The use of PlantUML supports automated parsing, validation, and aggregation of diagrams into view-specific artifacts.

\textbf{Prompt Design}. The \textit{Modeler} agent employs structured prompts for both generation and refinement. \texttt{Prompt~3} encodes architectural design knowledge (e.g., QA analysis, architectural patterns, and tactics) and enforces strict output constraints. Specifically, it requires to (1) derive architectural decisions from FRs, NFRs, and ASRs, (2) explicitly reason about QAs and trade-offs, and (3) produce exactly ten categories of standardized UML diagrams in PlantUML syntax aligned with the ``4+1'' view models (one Scenario view, three Logical views (Class, Object, State), three Process views (Activity, Sequence, Collaboration), two Development views (Package, Component), and one Physical view (Deployment)). The output format is strictly constrained to ensure compatibility with downstream processing. \texttt{Prompt~4} enables iterative improvement based on structured feedback from the \textit{Evaluator} agent. This prompt enforces a \textit{patch-based editing strategy}, requiring modifications only to affected diagram fragments, thereby preserving previously validated content and preventing unintended inconsistencies in other parts of the architecture model.

\begin{promptbox}[box:prompt3]{Simplified Prompt of the \textit{Modeler} Agent for Architecture Generation}
\textbf{Role}: Software architect.\\
\textbf{Input}: Functional Requirements (FRs); Non-Functional Requirements (NFRs); Architecturally Significant Requirements (ASRs); Domain Knowledge Reference; Past Architectural Decisions (Semantic Memory).\\
\textbf{Task}: 1. Summarize the proposed software architecture; 2. Analyze quality attributes and architectural trade-offs; 3. Recommend suitable architectural patterns and tactics; 4. Generate PlantUML diagrams based on the ``4+1'' architectural view model.\\
\textbf{Architecture Analysis}: Identify key quality attributes (e.g., scalability, security, maintainability) and explain how they influence architectural decisions, risks, and trade-offs.\\
\textbf{Architecture Design}: Select appropriate architectural patterns (e.g., Microservices, Layered), tactics, and justify them using FRs, NFRs, and ASRs.\\
\textbf{Diagram Generation}: Generate consistent PlantUML diagrams for the 4+1 architectural views.\\
\textbf{Diagram Rules}: Maintain consistent naming across diagrams; represent constraints and quality concerns; include architectural annotations, stereotypes, and deployment details where necessary.\\
\textbf{Output}: Produce structured architectural analysis and PlantUML syntax organized by the 4+1 architectural views.
\end{promptbox}

\begin{promptbox}[box:prompt4]{Simplified Prompt of the \textit{Modeler} Agent for Architecture Refinement}
\textbf{Role}: Software architect.\\
\textbf{Input}: Requirements Document (SRS); Existing UML Diagrams; Evaluator Feedback; Reference Knowledge; Past Modeling Lessons (Semantic Memory).\\
\textbf{Task}: Revise previously generated UML diagrams according to Evaluator feedback while preserving overall consistency and correctness.\\
\textbf{Core Rules}: 1. Treat the task as a patch operation rather than regeneration; 2. Only modify diagrams explicitly referenced in Evaluator feedback; 3. Preserve all unchanged PlantUML blocks; 4. Maintain naming, IDs, syntax validity, and cross-diagram consistency.\\
\textbf{Modification Rules}: 1. Apply minimal localized edits based on the \textit{Evaluator} feedback; 2. Add required 3-line analysis plans and compact notes only to modified diagrams; 3. Record assumptions when ambiguity exists; 4. Use conservative fixes for syntax/parsing issues.\\
\textbf{Consistency Rules}: 1. Preserve original diagram ordering and formatting; 2. Ensure valid PlantUML syntax; 3. Keep unchanged diagrams identical to the original input; 4. Abort modifications and return original diagrams unchanged if unauthorized edits are detected.\\
\textbf{Output}: Generate one Markdown file containing all UML diagrams, with only evaluator-authorized modifications applied.
\end{promptbox}

\textbf{Iterative Refinement and Memory Integration}. The \textit{Modeler} agent operates iteratively in coordination with the \textit{Evaluator} agent. \rev{Once the initial UML diagrams are generated, the \textit{Evaluator} immediately audits syntactic validity, cross-diagram consistency, and alignment with QA/ASR constraints (while system-level QA scenario generation and trade-off walkthroughs are subsequently synthesized in \texttt{Prompt~8}; see Section~\ref{sec:EvaluatorAgent}).} Based on Evaluator feedback (e.g., incompleteness or requirements misalignment), \rev{\texttt{Prompt~4}} incrementally refines the architecture by updating only the relevant parts\rev{, running autonomously without human intervention between rounds until passing validation or reaching a cap of three refinement iterations. Any ambiguities that cannot be resolved from the SRS alone are logged as explicit assumptions (\texttt{A\#}) inside diagram notes for subsequent human review.} This fine-grained revision mechanism reduces unintended modifications to previously validated architectural elements and improves traceability across iterations.

Throughout this process, the agent maintains episodic memory that records modeling iterations, feedback, and revisions, and leverages semantic memory to retrieve prior architectural knowledge and experience. This integration of \textit{episodic} and \textit{semantic memory} enables continuous improvement in requirements alignment, diagram consistency, and architectural completeness, producing a cohesive architecture model that serves as a reliable foundation for subsequent design stages.

\subsubsection{Designer Agent}\label{sec:DesignerAgent}
The \textit{Designer} agent synthesizes a draft architecture document by integrating analyzed requirements artifacts from the \textit{Analyst} agent and the multi-view architectural models produced by the \textit{Modeler} agent. It bridges high-level architectural abstractions and implementable system design by producing a detailed, developer- and operations-oriented documentation, including interface definitions, deployment configurations, data schemas, and traceability mappings, all stored in the shared artifact pool for subsequent validation and iteration.

Given these inputs, the \textit{Designer} agent constructs a coherent architecture description aligned with the ``4+1'' view model and incrementally refines it into detailed technical specifications, such as component responsibilities, technology selections, and interface contracts. This agent \rev{supports traceability by explicitly linking} architectural elements and design decisions to corresponding requirements. It further specifies operational aspects such as deployment configurations, scalability strategies, and operational configurations. Quality attributes (e.g., security and reliability) are systematically incorporated based on both requirement constraints and retrieved architectural knowledge. The final output is a structured architecture documentation accompanied by \rev{artifacts intended to inform subsequent development and operations planning after review.}

\textbf{Prompt Design}. The \textit{Designer} agent employs structured prompts for both generation and refinement. \texttt{Prompt~5} organizes the output into predefined sections (\texttt{Sections~A--L}), such as executive summary, traceability, architecture overview, detailed technical design, deployment, observability, testing, and trade-off analysis. To ensure \rev{that the generated architecture documentation is coherent, usable, and explicitly linked to architectural decisions}, the prompt enforces multiple constraints: (1) \rev{a structured traceability-and-rationale matrix (\texttt{Section~B}) explicitly binds each FR/NFR/ASR identifier to the corresponding UML diagram elements (\texttt{title:IDs}), components, concrete deliverable files (e.g., \texttt{openapi.yaml}, internal contracts, SQL DDLs, deployment manifests), and decision rationale;} (2) technology choices must \rev{compare multiple technology options through explicit trade-offs and justify the selected stack against the relevant requirements (\texttt{Sections~D} and~\texttt{J})~\cite{Kruchten2009DecisionView}}; (3) interfaces, data models, and deployment plans must be expressed as syntactically valid configurations \rev{that reference consistent UML element IDs without re-generating PlantUML source}; and (4) \rev{an end-of-document acceptance checklist verifies cross-section completeness.} \rev{Whereas the \textit{Modeler} addresses quality attributes structurally through patterns and tactics in UML views, \texttt{Prompt~5} translates those QAs and architectural choices into concrete API, schema, security, and operational specifications: the qualities named in \texttt{Prompt~5} (e.g., scalability, reliability, security, observability) establish baseline operational sections, while project-specific quality attributes are governed by the project's SRS and traced across \texttt{Sections~A, B, D,} and~\texttt{J}.} The prompt also incorporates external knowledge (see Section~\ref{sec:AgentSetting}) and prior design experience retrieved from \textit{semantic memory} to guide architectural decisions and promote reuse of validated architectural patterns. \texttt{Prompt~6} supports iterative improvement based on feedback from the \textit{Evaluator} agent. It strictly constrains modifications to only the artifacts identified in the Evaluator feedback and requires each change to be annotated with issue identifiers and rationale. This mechanism preserves previously validated content while enabling precise corrections.

\textbf{Iterative Refinement and Memory Mechanisms}. The \textit{Designer} agent operates iteratively in coordination with the \textit{Evaluator} agent. After each iteration, \rev{the \textit{Evaluator}'s documentation-review prompt audits FR/NFR/ASR traceability, cross-section alignment, and consistency with the \textit{Modeler}'s UML diagrams, and emits structured JSON issues.} The agent then performs targeted revisions via \texttt{Prompt~6} based on the feedback, updating only the affected sections to preserve previously validated architectural content across refinement iterations.

Throughout the process, the agent maintains \textit{episodic memory} to record inputs, generated artifacts, feedback, and revisions. It also distills reusable design insights (e.g., technology selection rationale and documentation practices) and stores them in \textit{semantic memory}. This memory integration of \textit{episodic} and \textit{semantic memory} enables continuous improvement and knowledge reuse across tasks. By combining structured prompt guidance, strict output constraints, iterative refinement, and memory-driven knowledge reuse, the \textit{Designer} agent produces internally consistent, requirement-traceable, and deployment-specifiable architecture documentation that faithfully reflects both system requirements and architectural intent.

\begin{promptbox}[box:prompt5]{Simplified Prompt of the \textit{Designer} Agent for Architectural Documentation Generation}
\textbf{Role}: Software design expert.\\
\textbf{Input}: Requirements Document (SRS); UML Diagrams; Reference Knowledge; Past Architectural Decisions (Semantic Memory).\\
\textbf{Task}: Design a production-ready architectural document aligned with requirements and UML diagrams, including detailed technical, operational, and deployment guidance for development teams.\\
\textbf{Core Requirements}: 1. Provide architecture overview, traceability, detailed subsystem design, APIs, data schemas, deployment, security, observability, testing, migration, and trade-off analysis; 2. Ensure every FR/NFR/ASR is traceable to architecture artifacts; 3. Include technology recommendations with justification linked to ASR/NFR IDs; 4. Generate architecture artifacts for both software development and system operation.\\
\textbf{Architectural Focus}: architectural patterns; deployment topology; API/interface contracts; scalability; reliability; security; observability; CI/CD; Kubernetes deployment; database consistency.\\
\textbf{Quality Rules}: use concise technical language; provide syntactically valid configuration/code snippets; specify technology version ranges; record assumptions and conflicts explicitly.\\
\textbf{Output}: Generate a complete architectural documentation in Markdown format.
\end{promptbox}

\begin{promptbox}[box:prompt6]{Simplified Prompt of the \textit{Designer} Agent for Architecture Refinement}
\textbf{Role}: Software design expert.\\
\textbf{Input}: Requirements Document (SRS); UML Diagrams; Evaluator Feedback; Reference Knowledge; Existing Architecture Document; Past Architectural Lessons (Semantic Memory).\\
\textbf{Task}: Refine the existing architecture documentation based on Evaluator feedback while preserving all unchanged content verbatim.\\
\textbf{Core Rules}: 1. Modify only sections or artifacts explicitly referenced in Evaluator feedback; 2. Preserve document structure and unchanged text exactly; 3. Apply minimal edits necessary to resolve issues; 4. Add inline notes for each fix and maintain a changelog/revision log.\\
\textbf{Technical Fix Requirements}: 1. Update or generate missing artifacts such as OpenAPI specs, internal contracts, SQL DDLs, Kubernetes manifests, or traceability matrices when flagged; 2. Ensure all added code/configuration snippets are syntactically valid; 3. Maintain consistency across APIs, deployment, data models, and architecture descriptions.\\
\textbf{Traceability and Validation}: 1. Ensure all FR/NFR/ASR requirements are traceable; 2. Re-run acceptance criteria verification after modifications; 3. Record unresolved conflicts or assumptions in the designated section.\\
\textbf{Output}: Generate a single updated Architecture Documentation in Markdown format.
\end{promptbox}

\subsubsection{Evaluator Agent}\label{sec:EvaluatorAgent}
The \textit{Evaluator} agent functions as the quality assurance and validation component within the MAAD framework. It systematically assesses requirements specifications, architectural models, and design documents to ensure requirements relevance, cross-view consistency, traceability completeness, and architecture-SRS alignment. By enforcing rigorous validation criteria and providing structured feedback, the \textit{Evaluator} agent supports continuous quality improvement of the output artifacts across all stages. Specifically, the evaluation process is guided by rule-based reasoning encoded in prompts and augmented by external reference knowledge. The \textit{Evaluator} agent performs structured checks, including requirement-to-architecture traceability, consistency, and validation that architectural tactics address the QAs specified in the SRS. At the architecture level, it further applies scenario-based analysis, risk identification, and trade-off assessment inspired by ATAM. It also derives quantitative indicators, such as mismatch counts and coverage metrics, to support metric-supported architecture quality assessment.

The \textit{Evaluator} agent operates at multiple stages in the workflow. First, it examines the requirements artifacts (FRs, NFRs, and ASRs) produced by the \textit{Analyst} agent, checking properties such as clarity, atomicity, measurability, and architectural relevance. Second, it assesses the architectural models generated by the \textit{Modeler} agent, verifying syntactic validity and semantic consistency of UML diagrams, and traceability to requirements. Third, it reviews the architecture documentation synthesized by the \textit{Designer} agent, ensuring technical completeness, internal coherence, and implementability. Finally, it conducts an evaluation by comparing the synthesized architecture against the original SRS, producing \rev{an ATAM-inspired preliminary architecture evaluation report} and a requirements-architecture mismatch analysis report.

\textbf{Prompt Design}. The \textit{Evaluator} agent employs task-specific prompts to standardize evaluation procedures and outputs. Requirement evaluation prompts enforce checklists that assess whether requirements are clearly defined, testable, internally consistent, and aligned with domain constraints and architectural objectives. The prompts further require structured outputs containing identified issues, affected requirement identifiers, severity levels, and recommended revisions. Design evaluation prompts incorporate validation rules for diagram parsing, cross-diagram consistency checks, and traceability analysis. For system-level evaluation, \texttt{Prompt~7} and \texttt{Prompt~8} generate a mismatch analysis report and \rev{an ATAM-inspired preliminary assessment}, both with standardized structures and explicit requirement references. These prompts also produce machine-readable outputs (e.g., JSON files) for downstream processing.

\begin{promptbox}[box:prompt7]{Simplified Prompt of the \textit{Evaluator} Agent for Architecture-Requirements Mismatch Analysis}
\textbf{Role}: Architecture evaluation expert.\\
\textbf{Input}: Extracted Requirements (FR/NFR/ASR); Architectural Documentation; UML Diagrams; Reference Knowledge.\\
\textbf{Task}: Analyze consistency and traceability between requirements and architecture; identify mismatches, omissions, conflicts, and architectural risks.\\
\textbf{Validation Rules}: 1. Every mismatch must reference requirement IDs and diagram/component IDs; 2. Missing or ambiguous requirements should be assigned inferred IDs; 3. Do not fabricate mismatches or evidence; 4. Prefer the original requirements descriptions when conflicts exist between requirements and diagrams.\\
\textbf{Mismatch Categories include}: missing traceability; inconsistent architecture design; unmet NFRs/ASRs; conflicting diagram elements; API/schema inconsistencies; security or scalability risks.\\
\textbf{Output}: Generate a structured mismatch analysis report, including traceability matrix, severity/risk assessment, remediation plan, verification mapping, and assumptions.
\end{promptbox}

\begin{promptbox}[box:prompt8]{Simplified Prompt of the \textit{Evaluator} Agent for Architecture Evaluation}
\textbf{Role}: Architecture evaluation expert.\\
\textbf{Input}: Requirements Document (SRS); UML Diagrams; Architectural Documentation; Reference Knowledge.\\
\textbf{Task}: 1. Analyze the architecture against business goals and quality attributes; 2. Evaluate architectural decisions using scenario-based analysis; 3. Identify risks, tradeoffs, sensitivity points, and mitigation strategies; 4. Generate traceability mappings between architectural decisions and requirements.\\
\textbf{Evaluation Focus}: Business drivers; quality attribute scenarios; architectural tactics and patterns; risk identification; tradeoff and sensitivity analysis; validation and remediation planning.\\
\textbf{Key ATAM Indicators include}: quality attribute tradeoffs; cross-cutting concerns; scalability and reliability constraints; security and performance risks; architecturally sensitive components and decisions.\\
\textbf{Output}: Generate a complete ATAM evaluation report, including risk registers, QA scenario analysis, tradeoff matrices, traceability mappings, and remediation plans.
\end{promptbox}

\textbf{Iterative Refinement and Memory Mechanisms}. 
The \textit{Evaluator} agent enables iterative refinement by generating structured feedback after each evaluation, including identified issues, severity levels, and actionable recommendations. Upstream agents (\textit{Analyst}, \textit{Modeler} and \textit{Designer}) use this feedback to revise their outputs, while the \textit{Evaluator} agent verifies whether previously identified architectural inconsistencies, requirement mismatches, and design-rule violations have been resolved in subsequent iterations.

To support this process, the \textit{Evaluator} agent maintains an episodic memory that records evaluation results, issues, and iteration outcomes, enabling traceability and preventing redundant reporting. In combination with the shared semantic memory, \rev{this accumulated knowledge supports consistent use of earlier feedback and helps avoid repeated issue reporting} across iterative evaluations.

\rev{To clarify the \textit{Evaluator}'s scope, its responsibilities separate into \textit{stage-level quality checks} and \textit{system-level cross-artifact analysis}. During generation, stage-level checks inspect a single agent's output (e.g., requirements, UML views, or documentation) and emit structured JSON feedback consumed by \texttt{Prompts~2, 4,} and~\texttt{6} for localized revision. Once all three upstream stages converge, system-level analysis runs across the assembled artifact set using external architectural standards and patterns as reference knowledge (Section~\ref{sec:AgentSetting}). Specifically, \texttt{Prompt~7} performs \textit{mismatch analysis} by cross-referencing extracted FR/NFR/ASR identifiers against UML elements and documented APIs/schemas to pinpoint unaddressed requirements, conflicting component responsibilities, or violated constraints. \texttt{Prompt~8} then synthesizes an \textit{ATAM-inspired preliminary report}~\cite{Kazman2000ATAM} deriving structured quality-attribute scenarios (specifying stimulus, environment, response, and measure) and scenario walkthroughs to identify sensitivity points, trade-offs, and risks; because a full ATAM evaluation requires interactive stakeholder workshops to prioritize business drivers and scenarios, this report serves as structured preparatory material for human architects rather than a final approval.}

\subsubsection{Knowledge-driven Agent Setting}\label{sec:AgentSetting}
In the MAAD framework, external knowledge is incorporated to enable context-aware generation of requirement-aligned and architecturally consistent artifacts. Software architecture design is inherently complex and cannot rely solely on input requirements and intermediate artifacts; it also requires established knowledge, including industry standards, architectural patterns, and domain-specific best practices. The external knowledge base consists of authoritative architectural standards and literature, including ISO/IEC/IEEE 42010:2022, 42020:2019 standards \cite{ISO_IEC_IEEE_42010_2022, ISO_IEC_IEEE_42020_2019}, as well as widely adopted architectural textbooks~\cite{Bass2021SAP, Martin2017CA, Richards2020FSA}. These sources are embedded into a vector database and retrieved during generation via Retrieval-Augmented Generation (RAG)~\cite{Lewis2020RAG}. The retrieved segments represent structured external knowledge, ensuring consistency with the knowledge representation used throughout the MAAD framework. Integrating such knowledge helps reduce information gaps, avoid overlooking critical considerations, and ensure that generated artifacts are grounded in both theory and practice. Knowledge infusion is primarily applied to the following two agents whose tasks depend most on the external context:

\textbf{Modeler Agent}: After receiving artifacts from the \textit{Analyst} agent, the \textit{Modeler} agent performs similarity searches over a vectorized knowledge base organized by architectural themes (e.g., layered architecture, component-and-connector patterns). The retrieved segments are incorporated into the prompt, enabling the agent to generate architecture views that are informed by both the requirements and relevant design knowledge.

\textbf{Designer Agent}: Before producing architecture documentation, the \textit{Designer} agent retrieves the most relevant segments from the same knowledge base via vector search. These segments provide background knowledge that supports design decisions and ensures alignment with established architectural practices.

Through pilot experiments, we select the top three most relevant segments for each query. This choice balances relevance and conciseness, providing sufficient contextual support without introducing unnecessary verbosity or noise. Overall, this knowledge-driven approach enhances decision-making and ensures that generated artifacts align with established practices, reducing omitted architectural considerations and improving decision consistency. \rev{Although both the external RAG knowledge base and the experience-derived semantic memory (see Section~\ref{sec:MemoryMechanism}) use embedding-based vector retrieval, they are maintained as two separate stores: the RAG knowledge base holds static architectural standards and textbook references, whereas the semantic memory accumulates reusable architectural guidance distilled from past design tasks.}

\subsection{Memory Mechanism Design}\label{sec:MemoryMechanism}
The memory design of the MAAD framework is inspired by the hierarchical memory mechanisms of human cognition, based on which we construct a three-layer memory architecture \cite{sumers2024cognitive}. The memory mechanism comprises \textit{working memory}, \textit{episodic memory}, and \textit{semantic memory}. Working memory functions as short-term memory, while episodic and semantic memory together form long-term memory. The overall workflow of the memory mechanism in MAAD is illustrated in Figure~\ref{F:Overview}.

\textbf{Short-Term Memory}: \textit{Working memory} represents the short-term memory layer of MAAD and maintains the most recent contextual information generated during task execution. It stores intermediate artifacts and interaction histories produced by the \textit{Analyst}, \textit{Modeler}, \textit{Designer}, and \textit{Evaluator} agents in each iteration. By preserving up-to-date reasoning context, working memory enables agents to reference prior steps and coordinate their actions effectively during collaborative problem solving. Specifically, working memory records the current revision round, the responsible agent, the artifacts generated in that round, the feedback provided by the \textit{Evaluator} agent, the identified aspects requiring modification, and the revised artifacts. This information supports the traceable refinement of architectural artifacts across design iterations and facilitates the progressive improvement of architectural design outputs (e.g., architectural views, documentation, and evaluation reports).

\textbf{Long-Term Memory}: The long-term memory in MAAD consists of two complementary components: \textit{episodic memory} and \textit{semantic memory}. 
\begin{itemize}
    \item \textit{Episodic memory} captures task-specific experiences accumulated during multi-agent collaboration. It stores artifacts, agent interactions, and key decisions across multiple iterations involving the \textit{Analyst}, \textit{Modeler}, \textit{Designer}, and \textit{Evaluator} agent. By preserving historical reasoning processes, episodic memory enables the system to revisit prior decisions, analyze outcomes, and support iterative improvement in subsequent cycles. Episodic memory organizes information at the granularity of individual tasks and their revision rounds. For each task, it records a task summary, the relevant agent responsible for generating the final artifacts, and the sequence of operations performed during each iteration. Additionally, it extracts and summarizes lessons learned from the task and assigns a confidence level. This mechanism supports task-level reflection and facilitates the initial extraction of reusable architectural patterns, design rationale, and evaluation insights from past executions. 
    \item \textit{Semantic memory} serves as the long-term knowledge repository of MAAD. In contrast to episodic memory, which focuses on task-specific experiences, semantic memory maintains generalized and reusable knowledge distilled from accumulated interactions and external sources. This knowledge includes architectural design principles, modeling rules, domain expertise, and recurring reasoning patterns. Semantic memory is constructed by aggregating and summarizing episodic memory after task completion. It extracts reusable insights, synthesizes architectural design rationale, and integrates them into a structured knowledge base for future reuse. Each semantic memory entry records the knowledge type, architectural principles and content, applicable conditions, related tasks and artifacts, and a textual summary to support vector-based retrieval.
\end{itemize}

Overall, the hierarchical memory mechanism enables filtering, aggregating, and abstracting architecture-design experiences and evaluation feedback, thereby supporting continuous knowledge accumulation and reuse in the architecture design process. It establishes a closed-loop knowledge engineering process within the MAAD framework, which enables an iterative architecture design process, integrating generation, evaluation, and refinement.

\rev{In Figure~\ref{F:Overview}, working memory operates along the inter-agent feedback loops to carry current intermediate artifacts and architecture review comments, whereas episodic and semantic memory reside in the left-hand knowledge panel. Implementation-wise, working memory is maintained as a structured runtime record for the active architecture design task; episodic memory stores per-stage architectural histories together with LLM-extracted design lessons in structured records; and semantic memory stores distilled, reusable architectural design principles in a shared vector database for embedding-based retrieval (separate from the external architecture-literature RAG knowledge base in Section~\ref{sec:AgentSetting}). Concretely, an \textit{architectural design experience} in episodic memory is a structured four-part record capturing (i) the requirement context, (ii) an initial design decision or model fragment, (iii) the \textit{Evaluator}'s critique, and (iv) the accepted revision. For instance, if an initial model lets a presentation controller access data stores directly, and the \textit{Evaluator} flags a security and modifiability violation against an access-control NFR, the resulting introduction of an intermediary service interface and repository contract is logged as one episodic experience; after task completion, its generalizable principle---isolating public-facing controllers from data stores under strict access-control constraints---is distilled into semantic memory with applicable conditions so future tasks retrieve context-relevant guidance rather than blindly copying past components.}

\section{Study Design}\label{sec:StudyDesign}
In this section, we present the Research Questions (RQs) in Section~\ref{sec:RQs} and provide an overview of the research process in Section~\ref{sec:Experiment Settings}.

\subsection{Research Questions}\label{sec:RQs}
% \yiran{Add some more ablation on things like evaluator feedback/memory could be better? Currently we only have ablation on the external knowledge.}

% \yiran{Also the cost/scalability discussion is missing.}

\begin{tcolorbox}[colback=gray!8, colframe=gray]
\textbf{RQ1: How effective is MAAD in automating the software architecture design?}
\end{tcolorbox}

\textbf{Rationale}: MAAD is a multi-agent framework for automating software architecture design. Although such automation shows significant promise, its practical effectiveness must be rigorously evaluated. Accordingly, RQ1 aims to assess the capability of MAAD to generate architectures that are structurally coherent, requirement-aligned, and practically acceptable to architects. To ensure a comprehensive evaluation, we first conduct a comparative analysis between MAAD and baseline approaches. We then present a case study where MAAD is applied to a realistic architecture design scenario to illustrate its strengths and limitations. Moreover, we perform a human evaluation involving industry experts to obtain qualitative insights into the quality of the generated architectures. Through this multifaceted evaluation, we aim to provide empirical evidence of MAAD's effectiveness, evaluate its performance, and validate its potential to improve efficiency and scalability in real-world software development.

% 结果：拿1个需求文档的case展示一下每个agent输出的artifacts（图）
% 对比MAAD跟metagpt（这种通用框架）生成的制品做对比
% 【改动】把RQ1的内容合并到Sec 3.2～3.5；RQ4的interview作为RQ1的一部分
% RQ1的结果展示：case study展示每个agent生成的制品；（2）对比MetaGPT的生成制品的结果；（3）interview的结果，证明MAAD框架的有效性，RQ4就取消掉。

\begin{tcolorbox}[colback=gray!8, colframe=gray]
\textbf{RQ2: To what extent does the infusion of external knowledge improve the quality of architecture design in the MAAD framework?}
\end{tcolorbox}
\textbf{Rationale}: MAAD incorporates external knowledge (e.g., existing system design, authoritative literature, and expert knowledge) to improve the architectural consistency and requirements alignment. Effectively integrating such knowledge is essential for enabling agents to make informed decisions that align with best practices and real-world constraints. RQ2 investigates how external knowledge can be infused and utilized within the MAAD framework, including techniques for knowledge extraction and representation. Addressing this RQ allows us to evaluate the contribution of external knowledge to improving the accuracy, consistency, and practical relevance of the generated architectural artifacts.

\begin{tcolorbox}[colback=gray!8, colframe=gray]
\textbf{RQ3: How do different LLMs affect the quality of architecture design in the MAAD framework?}
\end{tcolorbox}
\textbf{Rationale}: LLMs differ in their capabilities, domain knowledge, and reasoning patterns, which may influence the quality of the architectural design they produce. RQ3 examines the impact of using different LLMs within the MAAD framework. By analyzing their performance across various architecture design tasks, we aim to identify the strengths and limitations of each LLM. The answer to this RQ provides insights into selecting and configuring LLMs to achieve optimal performance and reliability in automated architecture design.

\begin{comment}
\begin{tcolorbox}[colback=gray!8, colframe=gray]
\textbf{RQ4: What are the main challenges encountered in deploying LLM-powered agents for automated architecture design?}
\end{tcolorbox}
\textbf{Rationale}: While LLM-powered agents offer great potential for automating architecture design, there are several challenges in their deployment. These challenges may include knowledge gaps, model biases, limited adaptability to new requirements, and issues related to the integration of agent generations into coherent architecture solutions. Additionally, practical deployment concerns such as scalability, real-time feedback, and maintaining the quality of generated design over time need to be addressed. RQ4 aims to identify these challenges and propose solutions or strategies to mitigate them through interviews with sophisticated architects. Understanding these obstacles is essential for refining the MAAD framework and ensuring its robustness and practical applicability in real-world software development environments.

% 结果：找架构师做个interviews，从practitioners的经验谈一下自动化架构设计面临的挑战
\end{comment}

\subsection{Experiment Settings}\label{sec:Experiment Settings}

\begin{figure}[hbtp]
    \centering
    \includegraphics[width=0.9\linewidth]{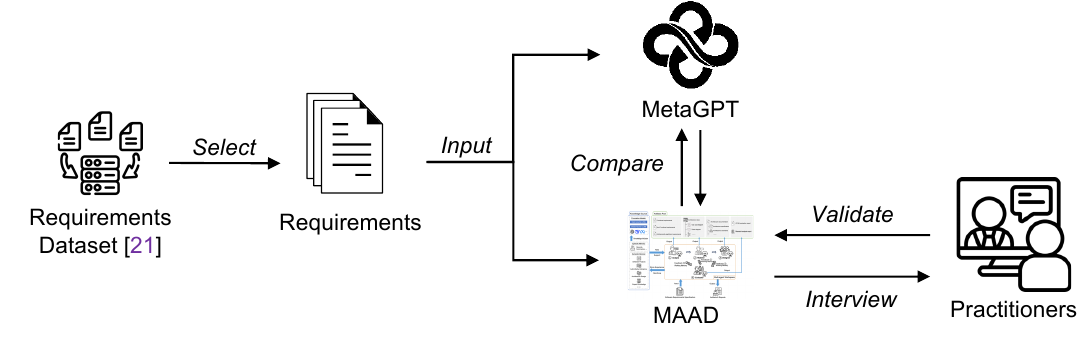}
    \caption{Process of the Study Design}\label{F:StudyDesign}
\end{figure}

This section describes the experimental design used to evaluate the MAAD framework from three complementary perspectives: effectiveness (RQ1), knowledge contribution (RQ2), and the impacts of LLMs' capability (RQ3). As illustrated in Figure~\ref{F:StudyDesign}, the study design process starts from requirements dataset selection, comparison between MAAD and a baseline, MetaGPT~\cite{hong2023metagpt}, and validation through practitioner feedback (i.e., professional software architects).

\rev{RQ1 evaluates MAAD with external-knowledge retrieval enabled. RQ2 contrasts this knowledge-enabled configuration with a configuration that removes retrieved external knowledge while retaining the same base task prompts and agent workflow. The added retrieved context is the intended difference between the conditions; disabling retrieval does not remove knowledge learned during the foundational LLM's pretraining.}

\subsubsection{Dataset}\label{sec:Dataset}
% \yiran{maybe a table here to briefly introduce the 10 projects?}
The requirements dataset used in this study is derived from Jin \textit{et al.} \cite{Jin2024erm}, which aggregates cyber-physical systems requirements from both public and industrial sources, including PURE~\cite{Ferrari2017Pure}, as well as proprietary industrial requirements documents~\cite{yang2022ipsf}. Our requirements dataset comprises 10 SRS cases, as shown in Table~\ref{T:dataset}. We use all 10 SRS cases for quantitative metric evaluation (see Section~\ref{sec:Metrics}), and select SFS (one of the 10 SRS cases) as a representative case for in-depth structural analysis, RQ2/RQ3 ablation, and qualitative validation. \rev{SFS is used as the running example because it is a self-contained interactive Web system with multiple actors, explicit educational workflow, and several quality concerns (e.g., usability, security, and feedback latency), which makes multi-view architecture modeling and documentation inspection tractable in a single case.}

\begin{table}[htbp]
  \centering
  \footnotesize
  \caption{Project Description of the Requirements Dataset}
  \label{T:dataset}
  \renewcommand{\arraystretch}{1.0}
  \begin{tabular}{p{0.7cm}p{12cm}}
    \toprule
    \textbf{Project} & \textbf{Description} \\
    \midrule
    C2C & An intelligent transportation infrastructure that interconnects regional traffic management centers to facilitate standardized traffic data exchange, incident reporting, and remote roadway device control. \\
    Case & A curated collection of classic, small-scale software engineering problem specifications and textbook case studies commonly used for requirements modeling, constraint analysis, and system design exercises. \\
    CCS & A real-time hardware control and monitoring system for the VLA radio telescope correlator, responsible for configuration management, autonomous fault recovery, and subsystem health tracking. \\
    CTS & A law enforcement information system designed to streamline crime investigation, prosecution workflows, and public-police interactions while enforcing strict role-based access and immutable audit controls. \\
    GCS & A comprehensive observatory control and data acquisition suite for the Gemini 8-meter telescopes, managing telescope operations, instrument scheduling, multi-user access, and remote observing protocols. \\
    HCS & A web-based home management system enabling remote monitoring and control of environmental parameters (temperature, humidity), security sensors, and appliance power states through wireless communication and a centralized gateway. \\
    LCS & A safety-critical traffic management system for controlling reversible highway lanes through coordinated operation of barrier gates, pop-up markers, and changeable message signs, featuring multi-layered safety interlocks and degraded-mode fallback capabilities. \\
    MEM & A ground data processing and archiving facility for the Mars Express mission's ASPERA-3 instrument, responsible for telemetry acquisition, IDFS data product generation, web-based public displays, and PDS-compliant archival submission. \\
    SFS & An interactive web-based educational game designed to improve sixth-grade students' fraction-solving skills through storyline-driven multiple-choice questions with adaptive feedback and administrator-configurable content. \\
    SSCS & A satellite attitude control system that performs sun acquisition and maintenance by processing gyroscope and sun sensor data to control thruster actuation across four operational modes: rate damping, pitch search, roll search, and sun cruise. \\
\bottomrule
  \end{tabular}
\end{table}

\subsubsection{Baseline}\label{sec:Baseline}
To assess the effectiveness of MAAD in comparison with existing multi-agent systems (MAS), we adopt MetaGPT \cite{hong2023metagpt} as the baseline. MetaGPT is a state-of-the-art MAS designed to generate software artifacts from a single requirements specification. From a structural perspective, MetaGPT comprises four agents \textit{Product Managers}, \textit{Architects}, \textit{Project Managers} and \textit{Engineers}. It simulates the workflow of a real-world software development organization by orchestrating these agents through well-defined Standard Operating Procedures (SOPs), thereby enabling end-to-end software development automation~\cite{hong2023metagpt}.

\rev{MetaGPT serves as a general-purpose software-development baseline, whereas MAAD explicitly targets architecture modeling, documentation, and evaluation. We compare overlapping artifact types to examine the effect of this architecture-oriented workflow. The comparison therefore distinguishes artifact-generation scope from the quality of architectural decisions; requesting more views alone does not establish better decisions.}
% \todo{CONFIG-01 / R1-6--7: build a run manifest for every figure and table: case, artifact path, model/provider/version, RAG setting, prompt hash, memory state, temperature, token limit, reasoning setting, retries, iteration count, and run ID. Resolve the RQ1 Qwen/RQ2 GPT and SFS/ten-case ambiguities using logs; leave unknown values explicit.}

\subsubsection{Metrics}\label{sec:Metrics}
% \yiran{1) no measurement for correctness: maybe we should have a manual validation or sth for this? 2) I think many metrics are designed for concrete code implementation rather than arch modules. Do we need some justification here? }
\rev{To describe selected structural properties of the generated artifacts, we use the following metrics:} Coupling Degree (CD) \cite{Chidamber1994}, Cohesion (Coh) \cite{Chidamber1994}, Interface Complexity (IC) \cite{kumari2011IC}, Structural Complexity (SC) \cite{McCabe1976}, State Complexity (StC) \cite{borger2000msm}, Component Coupling Density (CCD) \cite{lakos1996}, and State Machine Cyclomatic Complexity (SMCC) \cite{McCabe1976}. %Some metrics (e.g., $StC$ and $CCD$) are derived metrics grounded in established principles of state-based modeling \cite{McCabe1976}, coupling theory \cite{Chidamber1994}, and graph theory \cite{GraphTheory2008}, while others (e.g., $SMCC$) are based on well-established definitions \cite{McCabe1976}.

\rev{We select these metrics because they can be computed from the UML views MAAD generates and jointly characterize three structural perspectives across multiple diagram types: dependency and cohesion across component and class diagrams (CD, Coh, SC, CCD), interface surface across structural and interaction views (IC), and state-behavior complexity in state diagrams (StC, SMCC). We use these metrics to describe architectural designs alongside artifact inspection and practitioner feedback, not as a standalone architecture-quality score. The interpretation of these metrics depends on the abstraction level of the UML models, requirements coverage, and the rules used to extract structural elements from PlantUML; contextual fitness further requires evaluating the design against project requirements and stakeholder priorities. In practice, the reported values are obtained by extracting the relevant structural elements (e.g., dependencies, class/interface members, and state transitions) from the generated PlantUML sources and substituting them into the formulas below; the diagrams are not transformed into executable code for metric calculation.}

$CD$~\cite{Chidamber1994} measures the inter-module dependencies per module in the system, where $Dep_i$ represents the number of external dependencies of module $i$, and $N$ is the total number of modules (see Equation~\ref{eq:CD}). In architecture design, a lower CD typically indicates looser coupling and higher modularity, reducing the ripple effect of changes across components. 
\begin{equation}\label{eq:CD}
% 耦合度（Coupling Degree, CD），衡量模块之间依赖关系强度（适用于 Class / Component / Package Diagram）
% CD 越高 → 模块间耦合严重 → 架构脆弱
CD = \frac{\sum_{i=1}^{N} \mathit{Dep}_i}{N}
\end{equation}

$Coh$~\cite{Chidamber1994} quantifies the degree to which methods within a class share common attributes, %where $M_i$ denotes the methods of the class, and $\mathrm{disjoint}(M_i, M_j)$ indicates whether methods $M_i$ and $M_j$ access disjoint sets of attributes (see Equation~\ref{eq:Coh}). The metric ranges from 0 to 1, with values closer to 1 indicating higher cohesion and better encapsulation of related functionality within the class. 
where $M$ denotes the set of methods in a given class or component, and $|M|$ is its cardinality. $A(m_k)$ represents the set of attributes accessed by method $m_k \in M$. $I(\cdot)$ is the indicator function that returns $1$ if the intersection of attribute sets is empty (i.e., $A(m_i) \cap A(m_j) = \emptyset$, indicating disjoint attribute usage), and $0$ otherwise. The denominator $\binom{|M|}{2} = \frac{|M|(|M|-1)}{2}$ represents the total number of unique method pairs (see Equation~\ref{eq:Coh}). The metric ranges from $0$ to $1$, with values closer to $1$ indicating higher cohesion and stronger functional relatedness within the module.
\begin{equation}\label{eq:Coh}
% 内聚度（Cohesion），Coh越高，越接近 1 → 内聚越好；适用于class diagram
% Coh = 1 - \frac{\sum_{i \neq j} \mathrm{disjoint}(M_i, M_j)}{\binom{M}{2}}
Coh = 1 - \frac{\displaystyle\sum_{i=1}^{|M|} \sum_{j=i+1}^{|M|} I\big(A(m_i) \cap A(m_j) = \emptyset\big)}{\binom{|M|}{2}}
\end{equation}

$IC$~\cite{kumari2011IC} measures the average number of operations per interface in the system, where $\mathit{InterfaceOps}_i$ represents the number of operations provided by interface $i$, and $N$ is the total number of interfaces (see Equation~\ref{eq:IC}). Higher $IC$ values suggest more complex interfaces, which may increase the cognitive load for developers and raise the cost of system understanding and maintenance. 
\begin{equation}\label{eq:IC}
% 接口复杂度（Interface Complexity, IC），接口越复杂，维护成本越高；适用于component/sequence diagram
IC = \frac{\sum_{i=1}^{N} \mathit{InterfaceOps}_i}{N}
\end{equation}

$SC_{\mathit{norm}}$~\cite{McCabe1976} quantifies the complexity of software architecture based on graph theory, where $|V|$ represents the number of components (vertices) and $|E|$ denotes the number of dependency relationships (edges). The normalized form expresses the actual number of dependencies as a proportion of the maximum possible dependencies in a complete graph, expressing the edge density for the specified graph representation (see Equation~\ref{eq:SC}). Higher values indicate denser connectivity and potentially a more complex system structure. 
\begin{equation}\label{eq:SC}
% 结构复杂度（Structural Complexity），基于图论，V：组件数，E：依赖变数，适用于component diagram
SC_{norm} = \frac{|E|}{|V|(|V| - 1)}
\end{equation}

$StC$~\cite{borger2000msm} measures the overall complexity of a state diagram by summing the number of states and transitions. This metric captures both the size of the state space and the complexity of state interactions (see Equation~\ref{eq:StC}). Higher $StC$ values indicate more complex state machine behavior, which may increase the difficulty of behavioral verification and model maintenance. 
\begin{equation}\label{eq:StC}
% 状态复杂度（State Complexity， StC），适用于State diagram
StC = |\mathit{States}| + |\mathit{Transitions}|
\end{equation}

$CCD$~\cite{lakos1996} quantifies the average coupling per component in the software architecture, where $D_{in}(C_i)$ and $D_{out}(C_i)$ represent the number of incoming and outgoing dependency edges of component $C_i$, respectively, and $N$ is the total number of components (see Equation~\ref{eq:CCD}). This metric provides a comprehensive view of component interconnectivity by considering both dependencies imposed on a component and dependencies initiated by it. Higher $CCD$ values suggest greater architectural coupling, potentially impacting system modularity and evolvability. 
\begin{equation}\label{eq:CCD}
% 组件耦合密度 (Component Coupling Density, CCD)，N：component/package的总数，Din是指向Ci的依赖边数，Dout是从Ci指出的依赖边数
CCD = \frac{\sum_{i=1}^{N} (D_{in}(C_i) + D_{out}(C_i))}{N}
\end{equation}

$SMCC$~\cite{McCabe1976} adapts McCabe's cyclomatic complexity metric to state diagrams, where $E$ represents the number of transitions (edges), $N$ denotes the number of states (nodes), and $P$ is the number of connected components (typically $P=1$ for a single state machine) (see Equation~\ref{eq:SMCC}). This metric measures the number of linearly independent paths through the state machine, providing an indication of testing effort and structural complexity. Higher $SMCC$ values suggest more complex control flow and increased testing requirements to achieve comprehensive path coverage.
\begin{equation}\label{eq:SMCC}
% 状态机循环复杂度 (State Machine Cyclomatic Complexity, SMCC)，适用于State diagram
SMCC(G) = E - N + 2P
\end{equation}
% \yiran{maybe put each metric with its own description rather than put all metrics together? kind of too many metrics and hard to follow.}

\subsubsection{Selection of LLMs}\label{sec:LLMSelection}
To ensure a comprehensive evaluation across diverse model configurations and capabilities, we select four representative LLMs as foundational backbones for MAAD~\rev{spanning both proprietary and open-sourced ecosystems, parameter scales, and reasoning paradigms. Evaluating both closed-source commercial APIs and open-weight models examines how backbone capability affects artifact quality under the same workflow, while reflecting practical deployment needs where organizations choose between frontier cloud APIs and locally hostable open-weight models for confidentiality-sensitive requirements and architecture assets. We set the decoding temperature to $0.0$ for all four foundational LLMs.}

\begin{itemize}
\item \textbf{GPT-5.2}: OpenAI's proprietary multimodal LLM (parameters undisclosed) released on December 11, 2025~\cite{GPT5.2}. It extends GPT-4-class capabilities with improved long-context reasoning, multimodal processing, and tool-use support, targeting complex, real-time, and agent-oriented applications.

\item \textbf{Qwen 3.5}: Alibaba's open-weight multimodal LLM (Qwen3.5-397B-A17B) based on a Mixture-of-Experts architecture (397B total, 17B active parameters)~\cite{qwen35blog}. It supports long-context processing, multilingual generation, and multimodal inputs, and achieves strong performance in reasoning, coding, and vision-language tasks with efficient inference.

\item \textbf{Llama 3.3}: Meta's open-weight LLM (70 billion parameters)~\cite{dubey2024llama}, offering instruction-following, multilingual generation, and coding performance in a dense mid-scale architecture widely adopted for on-premise deployment.

\item \textbf{DeepSeek-R1}: DeepSeek's open-weight Mixture-of-Experts reasoning model (671 billion total parameters)~\cite{deepseekai2025}, trained via reinforcement learning to strengthen multi-step analytical reasoning on complex technical and software engineering tasks.

\end{itemize}

\subsubsection{Interviews}\label{sec:interview_design}
To further validate the practical utility of MAAD, we recruited six experienced software practitioners to solicit their feedback.

\rev{\textbf{Participant background.} Recruitment was purposive: we invited practitioners with substantial experience in software architecture design or architectural review, rather than requiring a single formal job title. Table~\ref{T:interview_participants} summarizes the anonymized profiles of the participants. Professional experience ranges from 5 to 25 years across academia and industry; three participants combine industrial architecture experience with academic roles. Inclusion targeted experience with architectural artifacts (requirements, multi-view models, and design documentation); we did not treat a separate UML-modeler certification as an inclusion criterion. Employer names are omitted to preserve anonymity.}

\begin{table}[htbp]
  \centering
  \footnotesize
  \caption{\rev{Anonymized background of the six interview participants}}
  \label{T:interview_participants}
  \renewcommand{\arraystretch}{1.05}
  {\revcolor
  \begin{tabular}{lllll}
    \toprule
    \textbf{ID} & \textbf{Sector} & \textbf{Experience} & \textbf{Role} & \textbf{Country} \\
    \midrule
    P1 & Academia & 13 years & Professor (industry architecture and academic experience) & China \\
    P2 & Academia & 11 years & Associate Professor (industry architecture and academic experience) & China \\
    P3 & Industry & 5 years & Senior Developer & China \\
    P4 & Industry & 8 years & Senior Developer & China \\
    P5 & Academia & 27 years & Professor (industry architecture and academic experience) & Italy \\
    P6 & Industry & 9 years & Architect & Singapore \\
    \bottomrule
  \end{tabular}}
\end{table}

\textbf{Interview Protocol}: We design an interview protocol by following the guidelines for empirical studies in software engineering proposed by Wohlin \textit{et al.}~\cite{Wohlin2012ESE}. We conduct semi-structured interviews with 3 open-ended questions that are designed to elicit practitioners' perspectives on automated architecture design using the MAAD framework. The interview questions allow the participants to freely and openly express their experiences and insights on the generated artifacts by MAAD from 10 real-world SRSs. The interview procedure consists of three parts: first, participants receive a concise overview of the study's objectives and are asked to review the artifacts generated by the MAAD framework for the ten user requirements cases. Second, interviewees are asked demographic questions (e.g., role, years of professional experience). Third, the first author conducts the interviews, each of which lasted 35 to 50 minutes. With the interviewees' consent, we audio-record the interviews and fully transcribe them for an in-depth analysis. The interview protocol and open questions are available in our replication package~\cite{onlinepackage_TOSEM}.

\textbf{Data Analysis}: The first author conducts a qualitative analysis of the transcripts, with the fourth author independently reviewing all coded segments to ensure consistency and mitigate bias. The data analysis proceeds as follows: (1) Extracting data: Transcripts are carefully read to identify salient comments regarding MAAD's artifact quality; (2) Coding data: Initial codes are generated to categorize participants' views on MAAD's generated artifacts. These codes guide subsequent thematic analysis; (3) Examination: To ensure analytical rigor, the fourth author independently reviews the coded data and resolves any discrepancies through discussion with the first author.

\section{Results}\label{sec:Results}
\subsection{Results of RQ1}\label{sec:RQ1_Results}
To answer RQ1, we evaluate the effectiveness of MAAD in automating software architecture design through a complementary mixed-methods approach. We first conduct an in-depth qualitative analysis using the ``Space Fraction System'' (SFS) from our dataset (see Section~\ref{sec:Dataset}) as a representative case. SFS is an interactive Web-based educational platform designed for sixth-grade students, which gamifies fraction arithmetic by presenting exercises, providing immediate feedback, and tracking learning performance. This case provides a rich ground for demonstrating the completeness, multi-view architecture consistency, and traceability of the architectural artifacts generated by MAAD. Additionally, we perform a quantitative evaluation across all 10 SRS cases, measuring architecture-level structural metrics and benchmarking MAAD against established multi-agent baselines.

% \yiran{Maybe pay more attention to the quantitative analysis of the 10 projects? I feel the current version kind of overfocuses on the SFS case. I feel like the evaluation here is kind of separated into two parts with no connection, can we make it like 10+1(SFS) projects for quantitative, then followed by a detailed case study on SFS}

% \yiran{Is metagpt the only baseline? I think its actually kind of outdated, and not dedicated for arch design. So better if we can have some more baseline here...}
% -->We agree that MetaGPT is not architecture-specific and may appear somewhat outdated. However, it is currently one of the few representative and reproducible multi-agent SE baselines, which makes it suitable as a reference point.

\subsubsection{Generated Artifacts of the MAAD Framework}\label{sec:RQ1_p1} 
Given the SRS of SFS, MAAD automatically generates a set of architecture artifacts that \rev{cover the artifact types targeted by the workflow}, including requirement specifications, multi-view architectural models, detailed architecture documentation, and architecture evaluation reports. 

\textbf{Requirements Artifacts.} MAAD produces structured requirements artifacts consisting of functional requirements (FRs), non-functional requirements (NFRs), and architecturally significant requirements (ASRs). In the SFS case, the generated FRs capture core system functionalities, such as adaptive question presentation, answer validation, feedback provision, performance tracking, and administrative content management. The NFRs explicitly specify QAs, including usability, compatibility, security, and performance, while ASRs highlight architecture design-driving concerns such as Web-based deployment constraints, standards-based technology stack, data persistence strategies, and real-time interaction. \rev{The outputs present FRs, NFRs, and candidate ASRs in separate artifacts; architectural significance can apply to both FRs and NFRs.}

\textbf{Multi-view Architecture Models.} Based on the requirements artifacts, MAAD generates a complete set of architecture models following the ``4+1'' view model. For the SFS case, this includes 10 UML diagrams spanning five views: (i) a \textit{Scenario View} is modeled using Use Case diagrams representing user interactions; (ii) \textit{Logical View} (including Class, Object, and State diagrams) capturing system structure and functionality; (iii) \textit{Process View} (including Activity, Sequence, and Collaboration diagrams) describing the dynamic aspects and run time behavior of the system; (iv) \textit{Development View} (including Package and Component diagrams) representing module organization from the developers' perspective; and (v) \textit{Physical View} (including Deployment) specifying runtime infrastructure. These models collectively provide a consistent and multi-perspective representation of the software architecture.

\textbf{Architecture Documentation.} MAAD synthesizes the above artifacts into a detailed architecture design documentation. In the SFS case, the documentation comprises multiple architectural views defined via PlantUML diagrams, which specify component responsibilities, interface definitions, logical data schemas, and deployment configurations. Importantly, the generated architecture design maintains explicit traceability between requirements and architectural elements through diagram annotations and notes, making the proposed links available for review.

\textbf{Evaluation Reports.} In addition to design artifacts, MAAD produces evaluation outputs, including \rev{an ATAM-inspired preliminary assessment} and a mismatch analysis report. For the SFS case, the ATAM report identifies key trade-offs (e.g., between Privacy vs. Functionality or Security vs. Availability), \rev{and the mismatch report records no critical mismatches in this SFS output. These reports are intended to assist human architects in reviewing candidates and making decisions; they are not a final confirmation of architectural correctness.}

\rev{The SFS case illustrates the sequence of requirements artifacts, multi-view models, documentation, and preliminary review reports produced by MAAD. These outputs demonstrate the workflow\textquotesingle s artifact-generation scope. The reported artifacts are available through the replication package~\cite{onlinepackage_TOSEM}.}

\subsubsection{Comparative Evaluation between MAAD and MetaGPT}\label{sec:RQ1_p2}
To evaluate the effectiveness of MAAD in automated software architecture design, we conducted a comparative analysis against MetaGPT~\cite{hong2023metagpt}, a multi-agent software development framework (see Section~\ref{sec:Baseline}). Although MetaGPT is not specifically designed for architecture design, it provides a baseline due to its end-to-end artifact generation capability through role-based agent collaboration. Given the differences in design objectives between MAAD and MetaGPT, we focused our comparison on overlapping artifact types generated by the two frameworks, including requirements analysis artifacts, architectural models, and design documentation. This enables meaningful evaluations of the two frameworks under comparable outputs.

\textbf{(1) Comparison of Requirements Analysis Artifacts}. In the requirements analysis phase, MetaGPT’s Product Manager agent produces a structured SRS, including: 

\begin{itemize}
    \item \textit{Product Goals} define a concise statement of the system’s primary features.
    \item \textit{User Stories} describe user usage scenarios and interaction processes.
    \item \textit{Competitive Analysis} compares competitors' features of similar products, highlights their strengths and weaknesses, and provides recommendations for feature optimization.
    \item \textit{Requirement Analysis} refines functional and non-functional requirements (such as performance and compatibility).
    \item \textit{Requirements Pool} includes a prioritized list of requirements.
    \item \textit{UI Design Draft} presents sketch layouts of UI design and shows basic interface design specifications.
\end{itemize}

Given the characteristics of MetaGPT, which ``\textit{takes a one-line requirement as input and outputs user stories, competitive analysis, requirements, data structures, APIs, documents, etc.}''\footnote{https://github.com/FoundationAgents/MetaGPT/}, it will automatically ``complete'' certain unreal details based on LLM's text generation capability. In other words, its generation process is inherently \textit{generative}, often introducing implicitly inferred requirements not directly grounded in the input SRS. For example, in the SFS case, MetaGPT autonomously enriches user behaviors (e.g., learning progress tracking) without explicit evidence from the input specification. More importantly, MetaGPT does not explicitly provide fine-grained requirements classification (i.e., FRs, NFRs, and ASRs), which is one of the disadvantages of common MAS approaches. Its output typically contains a small set of prioritized requirements, labeled as P0 (i.e., core functional requirements) or P1 (i.e., secondary functional requirements). Such outputs lack explicit distinctions among FRs, NFRs, and ASRs.

In contrast, the \textit{Analyst} agent in MAAD extracts structured and explicitly classified requirements artifacts from the SFS case, including i) 23 FRs with detailed descriptions, dependencies, and rationale, ii) 10 explicitly identified NFRs, and iii) 8 ASRs that directly inform architectural decisions. \rev{This SRS classification makes requirement categories and identifiers explicit,} \rev{providing explicit candidates for downstream modeling and review.}

\textbf{(2) Comparison of Architecture Modeling Capabilities}. For the system design of the SFS case, the Architect agent in MetaGPT generates a limited set of UML artifacts, including a class diagram and a sequence diagram (in Mermaid syntax), accompanied by brief textual descriptions (e.g., \texttt{Implementation Approach} description and \texttt{Anything UNCLEAR} declaration). While these artifacts capture partial structural and behavioral aspects, they lack multi-view architectural representation and do not explicitly address QAs or architectural trade-offs.

By contrast, the \textit{Modeler} MAAD agent systematically constructs architecture using the ``4+1''view models, generating multi-view, consistent UML artifacts. These views collectively capture structural, behavioral, developmental, and deployment perspectives, enabling a representation across multiple architectural views. Furthermore, MAAD enforces strict cross-view consistency and requirements traceability, supported by iterative evaluation from the \textit{Evaluator} agent. 

\begin{figure}[htbp]
    \centering
    % 第一个子图 (a)
    \begin{subfigure}[b]{\textwidth}
        \centering
        \includegraphics[width=\textwidth]{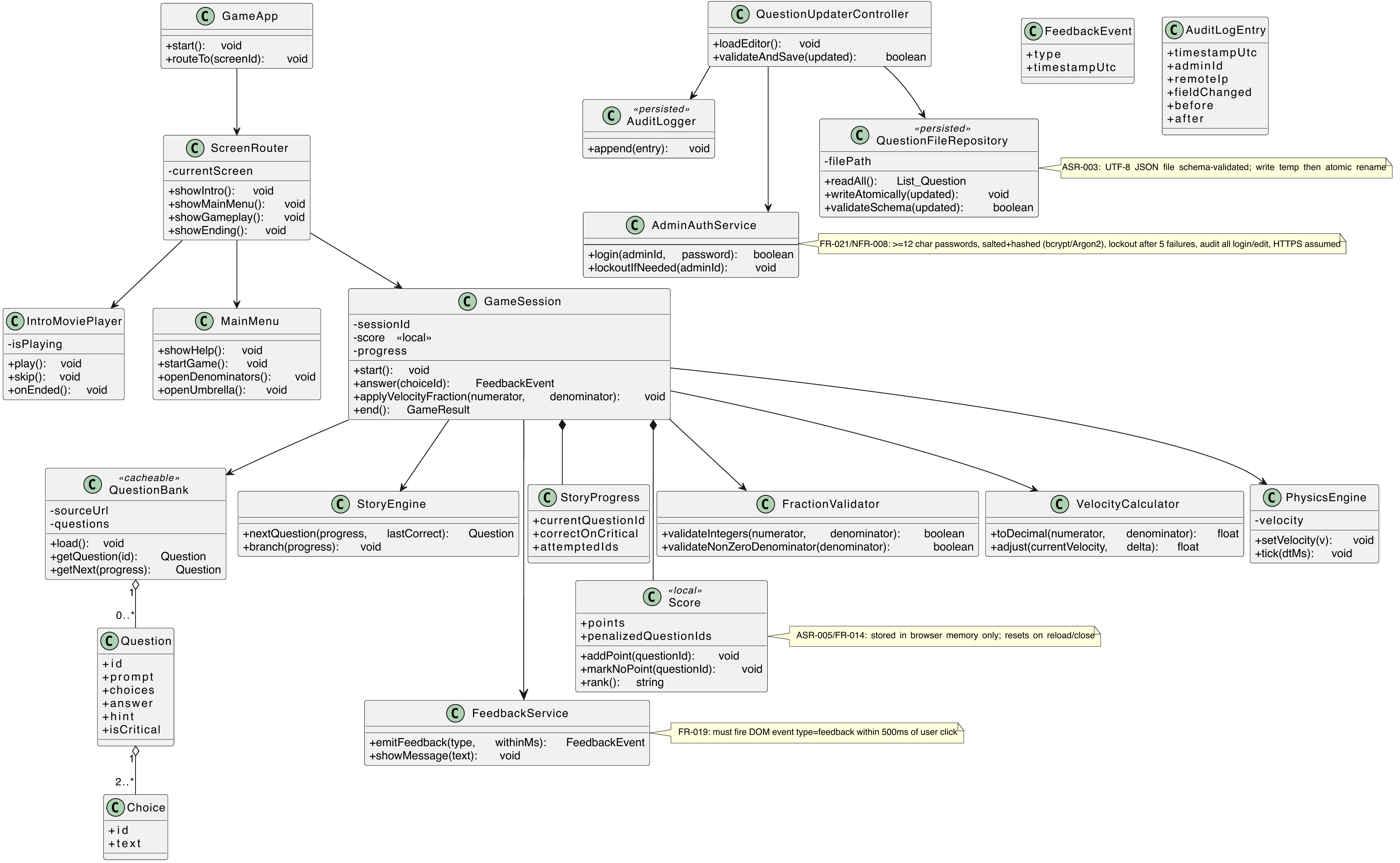}
        \caption{Class Diagram Generated by MAAD}
        \label{F:CDmaad}
    \end{subfigure}
    
    \vspace{0.25cm} % 控制上下子图之间的垂直间距（可按需调整）
    
    % 第二个子图 (b)
    \begin{subfigure}[b]{0.85\textwidth}
        \centering
        \includegraphics[width=\textwidth]{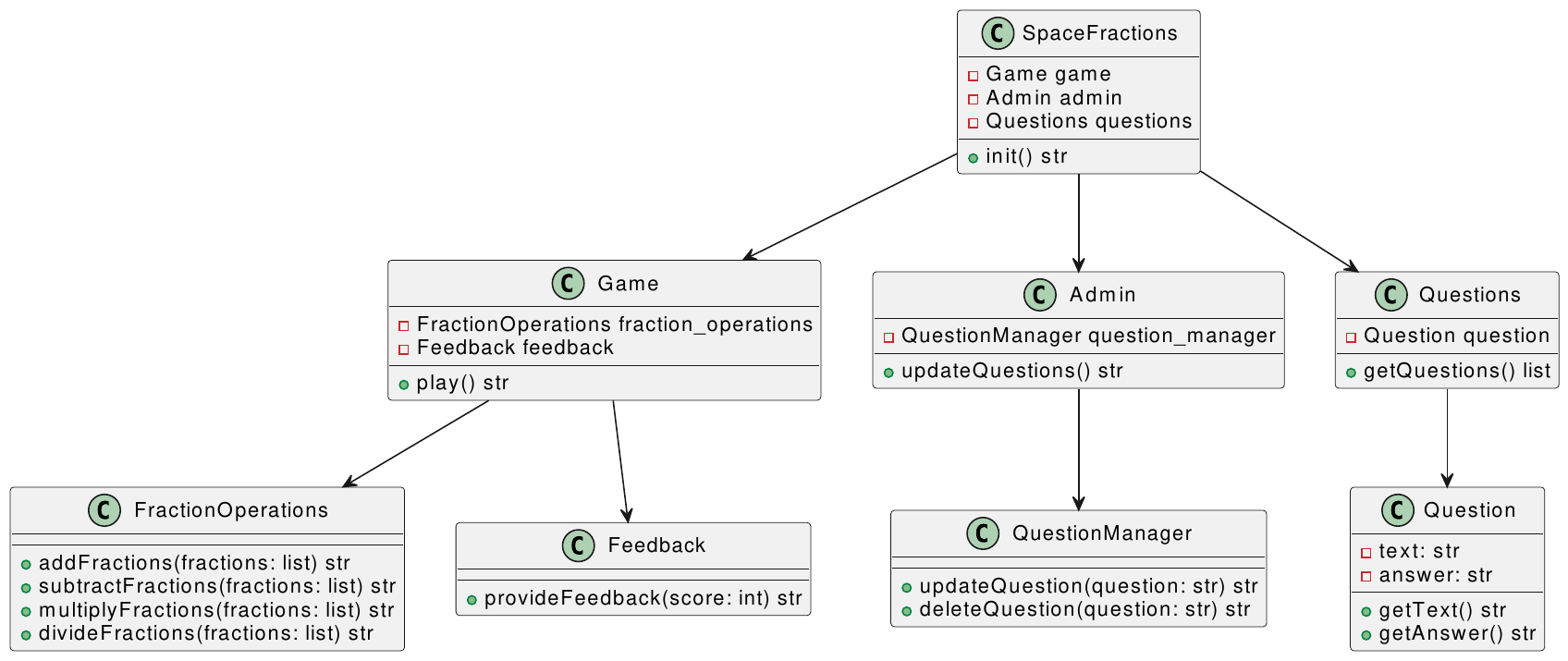}
        \caption{Class Diagram Generated by MetaGPT}
        \label{F:CDmetagpt}
    \end{subfigure}
    \caption{Comparison of the Class Diagrams Generated by MAAD and MetaGPT}
    \label{F:CDcomparison}
\end{figure}

\begin{figure}[htbp]
    \centering
    % 第一排：两个子图并列
    \begin{subfigure}[b]{0.67\textwidth}
        \centering
        \includegraphics[width=\textwidth]{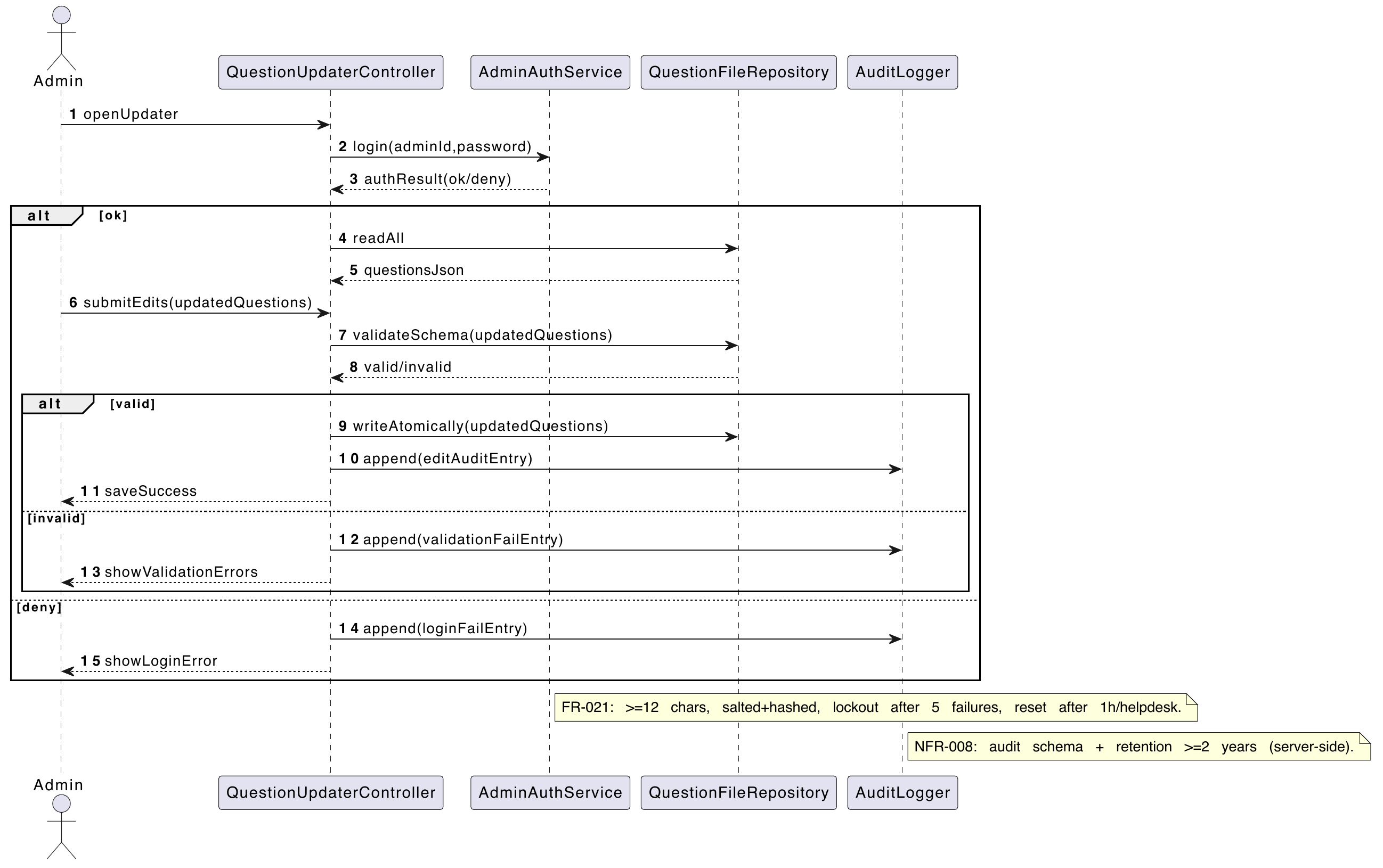}
        \caption{Sequence Diagram One Generated by MAAD}
        \label{F:SD1maad}
    \end{subfigure}%
    % \hspace{0.02\textwidth}
    
    \vspace{0.25cm} % 上下排垂直间距
    
    \begin{subfigure}[b]{0.53\textwidth}
        \centering
        \includegraphics[width=\textwidth]{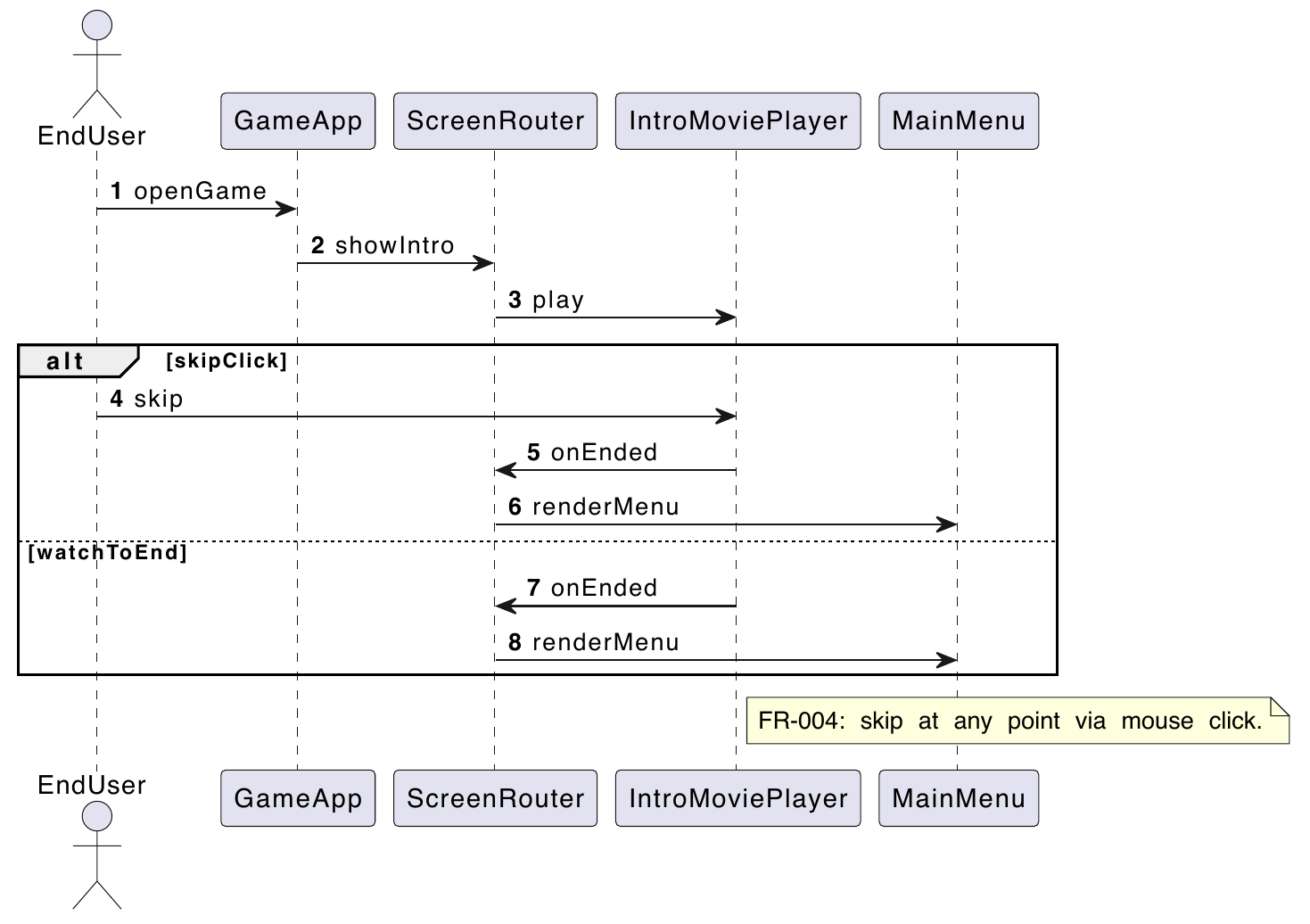}
        \caption{Sequence Diagram Two Generated by MAAD}
        \label{F:SD2maad}
    \end{subfigure}

    \vspace{0.25cm} % 上下排垂直间距

    % 第二排：第三个子图居中
    \begin{subfigure}[b]{0.48\textwidth}
        \centering
        \includegraphics[width=\textwidth]{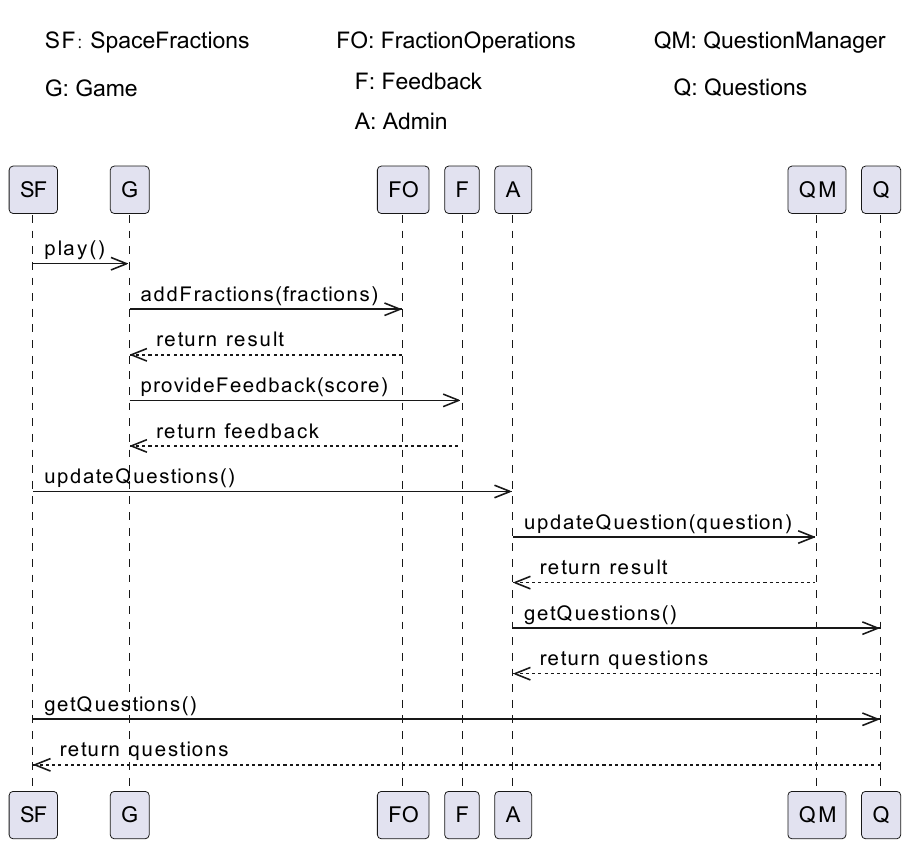}
        \caption{Sequence Diagram Generated by MetaGPT}
        \label{F:SDmetagpt}
    \end{subfigure}

    \caption{Comparison of Sequence Diagrams Generated by MAAD and MetaGPT}
    \label{F:SDcomparison}
\end{figure}

Figure~\ref{F:CDcomparison} and Figure~\ref{F:SDcomparison} compare the class and sequence diagrams generated by MAAD and MetaGPT, respectively. \rev{The class diagram generated by MAAD} (Figure~\ref{F:CDmaad}) \rev{contains more detailed classes and method descriptions} compared to the output from MetaGPT (Figure~\ref{F:CDmetagpt}). While MetaGPT produces a simplified, high-level abstraction with generic method signatures (e.g., returning \texttt{str} or \texttt{list}) and a flat hierarchy of FRs, MAAD captures a set of domain-specific entities, such as \texttt{PhysicsEngine}, \texttt{AuditLogger} and \texttt{QuestionFileRepository}, along with precise visibility modifiers and typed method signatures. Crucially, MAAD integrates requirements traceability directly into the structural design through annotations (e.g., ASR and NFR tags detailing password policies and storage constraints, respectively), thereby providing a much richer, implementation-oriented specification that minimizes ambiguity for downstream automated code generation. 

Regarding the sequence diagrams, MAAD generates multiple scenario-specific interaction diagrams for the SFS case study. Specifically, Figure~\ref{F:SD1maad} models the interactions involving the \texttt{Admin} actor for system management operations, and Figure~\ref{F:SD2maad} focuses on interactions involving the \texttt{EndUser} actor during quiz execution. These two sequence diagrams include more detailed interaction logic, such as fragments for conditional branching (e.g., handling validation success/failure, user skip actions). In contrast, MetaGPT (Figure~\ref{F:SDmetagpt}) produces a single, linearized sequence that conflates disparate concerns—mixing \texttt{Game} logic with administrative updates—and relies on abbreviated lifelines without capturing conditional paths, error handling, or return values. Furthermore, MAAD explicitly embeds requirements constraints (e.g., NFR-008) and precise method invocations directly into the interaction flow, providing a more detailed interaction description for scenario-based code generation. %, compared to MetaGPT's simplified, monolithic abstraction of the user scenario.

% Furthermore, to visualize these outputs, we employed PlantUML\footnote{https://plantuml.com/} to render the textual UML specifications as graphical diagrams. Notably, this visualization functionality is natively integrated into the MAAD framework, enabling automatic diagram generation directly from its output artifacts.

\textbf{(3) Comparison of Architecture Documentation}. In terms of documentation, MetaGPT outputs a JSON-based technical solution that includes the following fields: 

\begin{itemize}
    \item \textit{Required Python Packages} include the necessary Python packages.
    \item  \textit{Required Other Language Third-party Packages} refer to the selection of technology stack, which belongs to the technology dependency decision.
    \item \textit{Logic Analysis} defines the division of responsibilities of modules or files (e.g., \textsc{game.js} handles game logic, \textsc{admin.js} manages backend functions), reflecting the software architecture design.
    \item \textit{Task List} presents the code files to be implemented, which belong to the development task split.
    \item \textit{Full API Spec} describes API design specification.
    \item \textit{Shared Knowledge} describes the general design principles of the system (e.g., class or function sharing mechanisms), which belong to the architecture constraint description.
    \item \textit{Anything UNCLEAR} identifies issues that need to be clarified (such as browser compatibility, administrator interface design), which belong to requirements defect tracking and provide input for subsequent iterations.
\end{itemize}

Ideally, all the fields of the MetaGPT JSON files should contain content. However, our results show that several critical fields (e.g., \texttt{Required Python Packages} and \texttt{Full API Spec}) are often incomplete or missing, indicating limited stability in generating comprehensive architecture documentation. \rev{In contrast, the \textit{Designer} agent in MAAD produces architecture documentation covering decisions and rationale, component responsibilities, interfaces, deployment, and requirement-to-design links (see Section~\ref{sec:DesignerAgent}). These sections provide more detailed material for subsequent architecture review and development planning.}

\textbf{(4) Architecture Evaluation Outputs.} A key limitation of MetaGPT is the absence of a dedicated architecture evaluation mechanism. It does not assess the quality, consistency, and requirements alignment of the generated artifacts, making it difficult to ensure architectural validity. In contrast, MAAD integrates an \textit{Evaluator} that performs systematic, multi-stage evaluation and produces a mismatch report and an ATAM-inspired preliminary assessment, covering requirement alignment, cross-view checks, trade-offs, and risks.

\begin{table}[htbp]
     \centering
     \caption{Architecture-Level Metrics Comparison between MetaGPT and MAAD}
     \label{T:arch_metrics}
     \small
     \begin{tabular}{lcccc|cccc}
     \toprule
     \multirow{2}{*}{\textbf{Project}} 
     & \multicolumn{4}{c}{\textbf{MetaGPT}} 
     & \multicolumn{4}{c}{\textbf{MAAD}} \\
     \cline{2-9}
     & CD & Coh & IC & SC  
     & CD & Coh & IC & SC  \\
     \hline
     C2C  & 2.67 & 0.14 & 0.00 & 0.19  
          & 1.79 & 1.00 & 1.68 & 0.03  \\
     
     Case & 2.00 & 0.75 & 0.00 & 0.14  
          & 1.24 & 0.50 & 1.31 & 0.05  \\
     
     CCS  & 1.75 & 0.00 & 0.00 & 0.23 
          & 1.81 & 1.00 & 1.50 & 0.03  \\
     
     CTS  & 1.50 & 0.67 & 0.00 & 0.13  
          & 1.97 & 1.00 & 2.03 & 0.03  \\
     
     GCS  & 1.00 & 0.00 & 0.00 & 0.17  
          & 1.92 & 0.25 & 2.40 & 0.01  \\
     
     HCS  & 2.33 & 0.00 & 0.00 & 0.17 
          & 2.13 & 0.14 & 2.02 & 0.02  \\
     
     LCS  & 1.80 & 0.38 & 0.00 & 0.16 
          & 1.64 & 0.70 & 3.31 & 0.04  \\
     
     MES  & 2.00 & 0.25 & 0.00 & 0.14  
          & 1.57 & 1.00 & 1.83 & 0.03  \\
     
     SFS  & 1.33 & 0.40 & 0.00 & 0.20  
          & 1.60 & 0.28 & 1.35 & 0.03  \\
     
     SSCS & 2.33 & 1.00 & 0.00 & 0.17  
          & 2.00 & 1.00 & 3.33 & 0.02  \\\hline
     \textbf{Average} & 1.87 & 0.36 & 0.00 & 0.17 & 1.77 & 0.69 & 2.08 & 0.03\\
     \bottomrule
     \end{tabular}
     \end{table}

\textbf{(5) Quantitative Evaluation across Different Case Studies.} 
Table~\ref{T:arch_metrics} presents the quantitative comparison between MAAD and MetaGPT (using Qwen3.5 as the base LLM for the two frameworks) across 10 SRS cases from Jin \textit{et al}. \cite{Jin2024erm}. Due to the inherent differences between MAAD and MetaGPT (see Section~\ref{sec:RQ1_p2}), we restrict our comparison to their overlapping artifacts. For the common outputs, the evaluation applies four architecture-level metrics: Coupling Degree (CD) \cite{Chidamber1994}, Cohesion (Coh) \cite{Chidamber1994}, Interface Complexity (IC) \cite{kumari2011IC}, and Structural Complexity (SC) \cite{McCabe1976}. These metrics collectively evaluate modularity, internal consistency, interface design, and overall structure of the generated architectures.
% \yiran{I think we introduced 7 metrics previously but here we only keep 4. Either remove the 3 from the previous section or add them here? Or else we need some strong justification on why 1) they matter so we discussed them previously and 2) we have strong reasons to discard them here and 3) this will not damage the evaluation.}
% we had already mentioned it in Sec 5.1.2, we have added descriptions above.

As shown in Table~\ref{T:arch_metrics}, MAAD achieves lower CD than MetaGPT in the majority of cases, and higher or comparable cohesion in most projects. MetaGPT yields IC = 0 across all 10 projects, whereas MAAD reports explicit interface operations. MAAD also reports lower SC. Across the ten cases, the averages are CD = 1.77 versus 1.87, Coh = 0.69 versus 0.36, IC = 2.08 versus 0.00, and SC = 0.03 versus 0.17 for MAAD and MetaGPT, respectively. \rev{These differences show more explicit interfaces, higher cohesion, and lower structural complexity in the generated representations.}

Overall, the quantitative comparison results demonstrate that MAAD generates architectures with (1) lower structural complexity, (2) higher cohesion, and (3) explicitly defined interface interactions, while maintaining comparable or improved coupling control. These characteristics collectively indicate that MAAD produces more modular, structured, and architecturally sound design than MetaGPT.

\subsubsection{Practitioner-Based Qualitative Evaluation}\label{sec:RQ1_p3}
% \yiran{For the interview, 1) I think better if we have some score along with some comments? 2) Better if let them rate both MAAD and baseline Metagpt? 3) add some case findings to support the generalized claims in this section? 4) Any negative feedback? I think the result cannot be perfect. }
To complement the quantitative analysis and case study, we conducted semi-structured interviews with six experienced software practitioners (5--25 years of professional experience; see Table~\ref{T:interview_participants}) to qualitatively assess whether MAAD can effectively support real-world software architecture design activities. The evaluation assesses the practicability and real-world applicability of the generated architectural artifacts. Through thematic analysis of the interview transcripts, we identified four overarching evaluation dimensions: \textbf{(1)} perceived quality of generated artifacts, \textbf{(2)} practical utility in supporting architectural work, \textbf{(3)} adaptation considerations and human-in-the-loop, and \textbf{(4)} future enhancement.

\textbf{Perceived Architectural Quality.} 
Across all interview participants, MAAD-generated artifacts were consistently regarded as well-structured and aligned with established architectural principles. Participants noted that the outputs systematically reflect core design concepts such as modular decomposition, QAs, and traceability from requirements to architecture. The artifacts were considered comparable to early-stage architecture drafts produced in practice. In particular, one senior practitioner (P5) emphasized that the level of structural organization and the detail of architecture documentation align with standard industry expectations for initial design deliverables, and in certain respects, surpass them. \rev{These observations support the perceived usefulness of the outputs as initial design material.}

\textbf{Practical Utility and Knowledge Support.} 
All participants highlighted the practical usefulness of MAAD as a supportive tool in the architecture design process. A key strength mentioned is the ability to integrate and utilize external knowledge, which enables more comprehensive consideration of design alternatives and QAs. Participants particularly valued the automatically generated evaluation artifacts (e.g., mismatch reports), which help identify potential inconsistencies between requirements and generated architecture during architecture analysis. These features were perceived as beneficial for reducing manual architecture analysis and documentation efforts, and improving architecture validation efficiency. Furthermore, several participants (P1, P3, and P6) noted that the knowledge-driven mechanism enhances adaptability to domain-specific scenarios, making the MAAD framework applicable beyond generic system design tasks.

\textbf{Adaptation Considerations and Human-in-the-Loop.} 
While participants acknowledged the capability of MAAD in generating high-quality architectural artifacts and its potential to improve architecture design efficiency, they emphasized the importance of maintaining human involvement in the architecture design process. Rather than viewing MAAD as a fully autonomous replacement, practitioners consistently framed it as an intelligent assistant that augments human decision-making. Participants also noted that the effective deployment of MAAD would require adaptation to project-specific contexts and organizational practices. Software systems should support iterative refinement based on user feedback, enabling architects to adjust generated artifacts and progressively align recommendations with evolving requirements and domain constraints. In particular, participants highlighted that architectural design often involves contextual judgment, implicit assumptions, and domain-specific constraints that benefit from human expertise. As P6 suggested, human architects as gatekeepers were considered essential for ensuring accountability, interpretability, and alignment with project-specific requirements, especially in complex or safety-critical systems.

\textbf{Insights for Future Enhancement.} 
Participants also provided constructive suggestions for further improving the MAAD framework. One key direction is the specialization of agent capabilities through tailored or domain-specific LLMs, which could improve architecture-related reasoning in tasks such as domain-aware architectural decision-making and interface design. Another important aspect is the incorporation of stronger external tools and memory mechanisms, enabling the system to leverage existing tools and accumulated prior experience across projects. Additionally, participants emphasized the persistent challenge posed by tacit knowledge: ``\textit{During software architecture design, much tacit knowledge is hard to capture and cannot yet be leveraged, which remains a major obstacle in the automated design driven by knowledge}''.

Overall, the qualitative evaluation and findings indicate that \rev{MAAD is perceived as a practically useful assistant} for supporting software architecture design. Its strengths lie in generating structured, principle-aligned artifacts and leveraging knowledge to enhance architecture design quality. At the same time, practitioners regard MAAD as a complementary tool that can be strengthened through integration with human expertise, highlighting opportunities for further improvement in domain-specific specialization, cross-project reuse of architectural knowledge, and the transparency and interpretability of architecture evaluation results.

\begin{tcolorbox}
\rev{\textbf{RQ1 Summary:} \textit{MAAD automatically generates structured requirements artifacts, multi-view models, architecture documentation, and preliminary evaluation reports. Compared with the selected MetaGPT outputs, it provides more detailed architectural descriptions together with dedicated evaluation reports, with differences in the reported structural indicators. Practitioners regard the outputs as useful initial design deliverables and review aids. The evidence supports automatic artifact generation with human review and approval of architectural decisions.}}
\end{tcolorbox}

\subsection{Results of RQ2}\label{sec:RQ2_Results}
To investigate the impact of external knowledge infusion on architecture design quality (RQ2), we conduct a comparative study of MAAD under two configurations: (1) \textbf{with RAG}, where the \textit{Modeler} and \textit{Designer} agents leverage external architectural knowledge, and (2) \textbf{without RAG} (no-RAG), where architecture generation relies solely on the input SRS and internal reasoning. 
% The external knowledge base consists of authoritative architectural standards and literature, including ISO/IEC/IEEE 42010:2022, 42020:2019 standards \cite{ISO_IEC_IEEE_42010_2022, ISO_IEC_IEEE_42020_2019}, as well as widely adopted architectural textbooks~\cite{Bass2021SAP, Martin2017CA, Richards2020FSA}. These sources are embedded into a vector database and retrieved during generation via Retrieval-Augmented Generation (RAG)~\cite{Lewis2020RAG}. 
\rev{As stated in Section~\ref{sec:Experiment Settings}, the two configurations keep the same base task prompts and agent workflow; the intended difference is the presence or absence of retrieved external knowledge.} Following the study design (Section~\ref{sec:StudyDesign}), we evaluate the impact of knowledge infusion from two complementary perspectives: (1) \textbf{structural comparison} of architectural artifacts and (2) \textbf{quantitative comparison} using architecture-level metrics.

\begin{figure}
    \centering \includegraphics[width=1\linewidth]{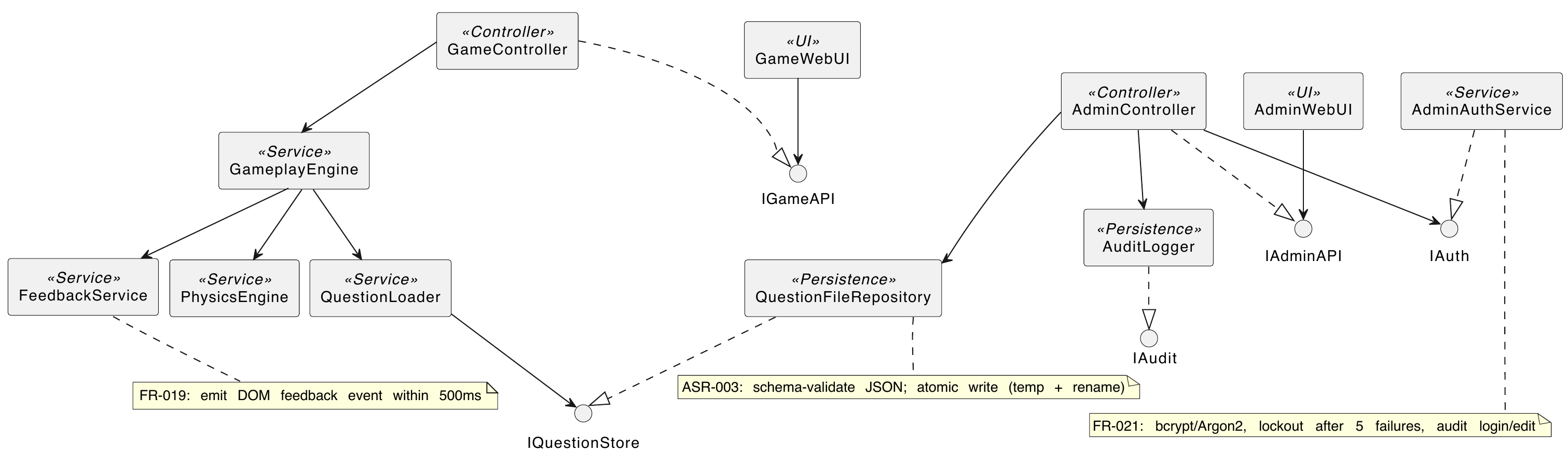}
    \caption{The Component Diagram of SFS Generated with Reference Knowledge}
    \label{fig:component_view_withRAG}
\end{figure}

\begin{figure}
    \centering\includegraphics[width=1\linewidth]{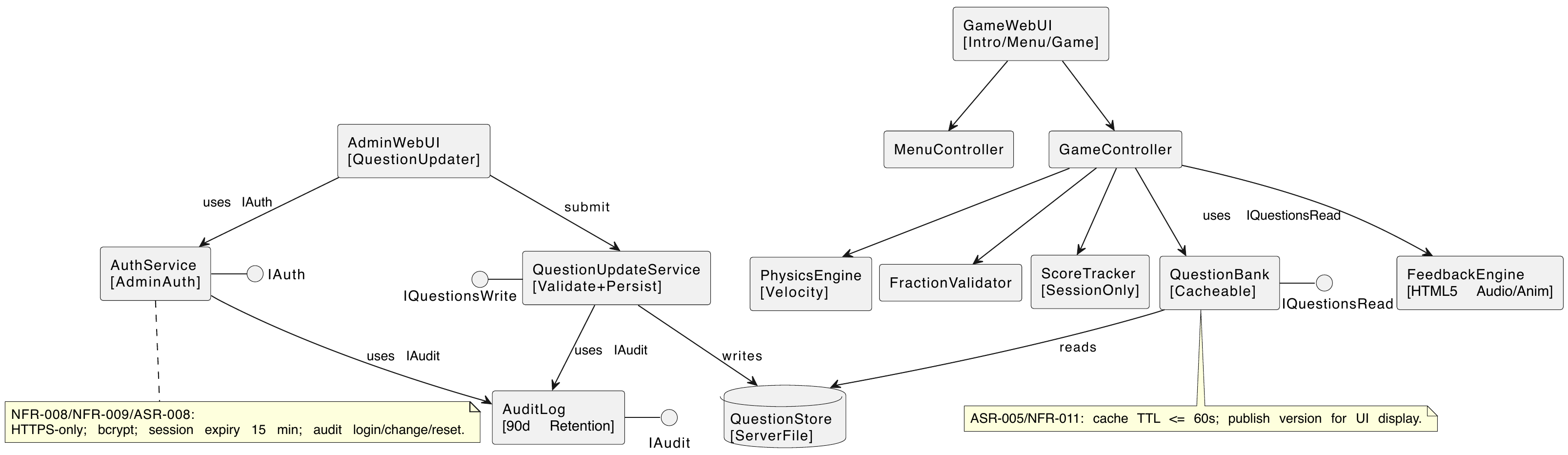}
    \caption{The Component Diagram of SFS Generated without Reference Knowledge}
    \label{fig:component_view_withoutRAG}
\end{figure}

\textbf{(1) Structural Comparison.} Using the Space Fraction System (SFS) as a representative case, the component views generated with and without external knowledge are shown in Figures~\ref{fig:component_view_withRAG} and~\ref{fig:component_view_withoutRAG}. \rev{The with-RAG view labels architectural roles and inserts an intermediate service layer.} It employs UML stereotypes (e.g., \texttt{<<Controller>>}, \texttt{<<Service>>}, \texttt{<<Persistence>>}, and \texttt{<<UI>>}) together with interfaces such as \texttt{IGameAPI} and \texttt{IQuestionStore}. \texttt{GameplayEngine} mediates between \texttt{GameController} and the lower-level services \texttt{FeedbackService}, \texttt{PhysicsEngine}, and \texttt{QuestionLoader}. Technical constraints are attached to the corresponding components, enabling direct requirement-to-component traceability.

In contrast, the no-RAG view \rev{uses a flatter topology and does not apply these role stereotypes.} \texttt{GameController} connects directly to lower-level components such as \texttt{PhysicsEngine}, \texttt{FractionValidator}, and \texttt{ScoreTracker}, so the diagram emphasizes functional dependencies. \rev{The same view still specifies interfaces, including \texttt{IQuestionsRead}, \texttt{IQuestionsWrite}, \texttt{IAuth}, and \texttt{IAudit}, and it attaches requirement annotations.} \rev{Validation and scoring remain separate elements (\texttt{FractionValidator} and \texttt{ScoreTracker}), while \texttt{QuestionUpdateService} carries a combined validate-and-persist label.}

The distinction is the level of abstraction. The RAG version emphasizes architectural roles through stereotypes and the intermediate \texttt{GameplayEngine}, while the no-RAG version records direct dependencies among concrete components together with interfaces and requirement annotations. These differences suggest that, in this case, retrieved architectural knowledge \rev{leads the agents to expose roles and layering in the component view}. \rev{Without that retrieved context, the same prompts and workflow still produce a component view, organized around direct component dependencies.}

\textbf{(2) Quantitative Comparison.} % 拿GPT5.2跑的
To further quantify the impact of external knowledge infusion, Table~\ref{tab:norag_vs_maad} compares MAAD with and without RAG across seven architecture-level metrics (see Section~\ref{sec:Metrics}). %\rev{The direction of change depends on the metric and the project.}

\begin{itemize}
\item \textbf{Coupling metrics show mixed effects.} Compared with MAAD without RAG, $CD$ decreases in 4 projects (e.g., HCS and C2C), increases in 5 others (e.g., GCS and MES), and remains unchanged in SFS. \rev{$CCD$ is lower in 6 of the 10 projects and higher in the other 4 (CCS, MES, SFS, and SSCS). Lower $CCD$ is the more frequent coupling change.}
\item \textbf{Cohesion decreases in most projects.} Compared with MAAD without RAG, $Coh$ decreases in 8 of the 10 projects, is unchanged in CCS, and increases only in LCS. \rev{As responsibilities are distributed across more specialized components, connections among elements within a component can decrease, which lowers this cohesion score.}
\item \textbf{Interface complexity varies by project.} $IC$ increases in some systems (e.g., Case and LCS) and decreases in others (e.g., SSCS and SFS).
\item \textbf{Structural complexity remains stable.} $SC$ stays within 0.01--0.03 under both configurations, with only minor increases in 4 projects. External knowledge does not substantially change this density measure.
\item \textbf{State complexity varies by project.} \rev{$StC$ and $SMCC$ exhibit notable variability. $StC$ is higher with RAG in GCS and MES and lower in Case and LCS. $SMCC$ is higher in GCS and Case and lower in LCS.}
\end{itemize}

\rev{Overall, the averages show that MAAD with RAG improves average component coupling density ($CCD$ from 2.78 to 2.54), while $CD$ and $IC$ barely change (from 2.16 to 2.13, and from 2.35 to 2.34). Cohesion is lower on average ($Coh$ from 0.64 to 0.50), consistent with distributing responsibilities across more specialized components. $StC$ decreases (from 85.50 to 78.80) while $SMCC$ increases (from 4.60 to 6.20). Knowledge infusion therefore yields a coupling-density gain together with a cohesion trade-off, rather than consistent improvements across all metrics.}

\begin{table*}[htbp]
\centering
\setlength\tabcolsep{3.5pt}% adjust column padding
\caption{Architecture-Level Metrics Comparison between MAAD (without RAG) and MAAD (with RAG)}
\label{tab:norag_vs_maad}
\small
\begin{tabular}{lccccccc|ccccccc}
\toprule
\multirow{2}{*}{\textbf{Project}}
& \multicolumn{7}{c}{\textbf{MAAD (without RAG)}}
& \multicolumn{7}{c}{\textbf{MAAD (with RAG)}} \\
\cline{2-15}
& CD & Coh & IC & SC & StC & CCD & SMCC
& CD & Coh & IC & SC & StC & CCD & SMCC \\
\hline
C2C  & 2.37 & 0.69 & 1.86 & 0.02 & 67 & 2.91 & 6
     & 1.93 & 0.61 & 2.41 & 0.02 & 74 & 2.29 & 4 \\
Case & 1.98 & 0.67 & 2.00 & 0.02 & 99 & 3.78 & 1
     & 2.13 & 0.45 & 3.48 & 0.03 & 52 & 2.29 & 6 \\
CCS  & 2.38 & 0.53 & 2.02 & 0.01 & 105 & 2.17 & 4
     & 2.16 & 0.53 & 1.73 & 0.01 & 88 & 2.36 & 3 \\
CTS  & 2.10 & 0.80 & 1.89 & 0.02 & 107 & 2.56 & 3
     & 2.27 & 0.77 & 1.96 & 0.03 & 54 & 2.44 & 4 \\
GCS  & 2.42 & 0.76 & 1.73 & 0.02 & 97 & 3.29 & 8
     & 2.63 & 0.56 & 1.97 & 0.02 & 141 & 2.76 & 16 \\
HCS  & 2.11 & 0.55 & 2.25 & 0.02 & 76 & 3.17 & 4
     & 1.57 & 0.40 & 1.90 & 0.02 & 69 & 2.20 & 2 \\
LCS  & 1.98 & 0.50 & 2.40 & 0.02 & 89 & 2.29 & 7
     & 2.19 & 0.64 & 3.83 & 0.03 & 50 & 1.78 & 3 \\
MES  & 1.97 & 0.80 & 2.62 & 0.01 & 84 & 3.00 & 7
     & 2.30 & 0.27 & 2.58 & 0.03 & 114 & 4.00 & 6 \\
SFS  & 2.05 & 0.47 & 2.74 & 0.02 & 73 & 2.17 & 1
     & 2.05 & 0.33 & 1.97 & 0.02 & 68 & 2.73 & 11 \\
SSCS & 2.25 & 0.67 & 3.95 & 0.03 & 58 & 2.44 & 5
     & 2.09 & 0.43 & 1.57 & 0.02 & 78 & 2.50 & 7 \\\hline
\textbf{Average} & 2.16 & 0.64 & 2.35 & 0.02 & 85.50 & 2.78 & 4.60 & 2.13 & 0.50 & 2.34 & 0.02 & 78.80 & 2.54 & 6.20 \\
\bottomrule
\end{tabular}
\end{table*}

\begin{tcolorbox}
\rev{\textbf{RQ2 Summary:} \textit{External knowledge changes the illustrated SFS component view: the with-RAG diagram makes role stereotypes and an intermediate service layer explicit, while the no-RAG diagram uses a flatter controller-to-service decomposition and still specifies interfaces and requirement annotations. Across the ten projects, the structural metrics do not get consistent improvements. Component coupling density is lower in 6 of 10 cases and cohesion is lower in 8 of 10, while interface and state-related metrics vary by project.}}
\end{tcolorbox}

\subsection{Results of RQ3}\label{sec:RQ3_Results}
\begin{figure}[htbp]
    \centering   % 第一排：两个子图并列
    \begin{subfigure}[b]{0.48\textwidth}
        \centering
        \includegraphics[width=\textwidth]{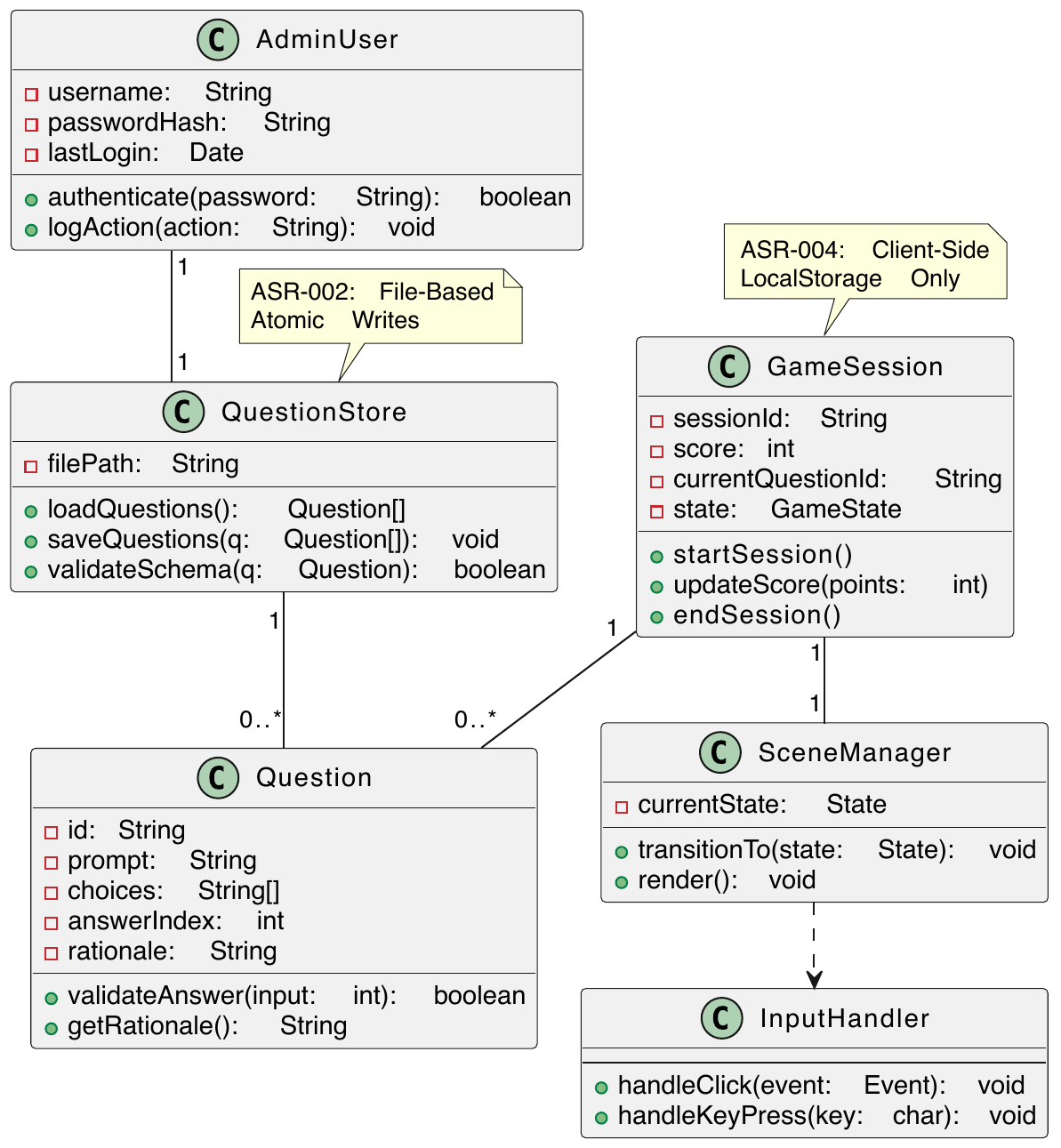}
        \caption{The Class Diagram of SFS Generated by Qwen3.5}
        \label{F:Qwen_classD}
    \end{subfigure}%
    \hspace{0.04\textwidth}%
    \begin{subfigure}[b]{0.46\textwidth}
        \centering
        \includegraphics[width=\textwidth]{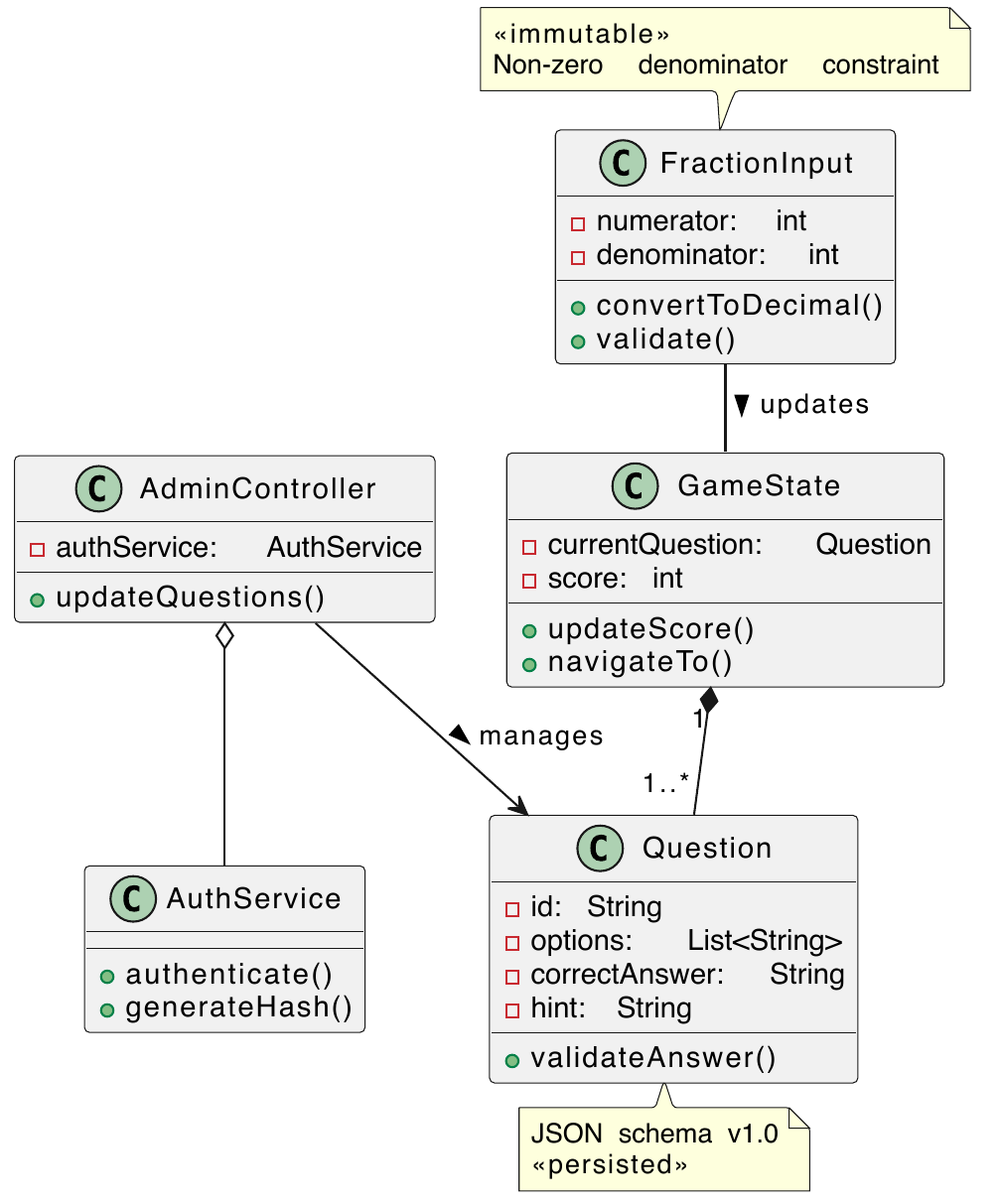}
        \caption{The Class Diagram of SFS Generated by DeepSeek-R1}
        \label{F:DeepSeek_ClassD}
    \end{subfigure}

    \vspace{0.25cm} % 上下排垂直间距

    % 第二排：第三个子图居中
    \begin{subfigure}[b]{0.48\textwidth}
        \centering
        \includegraphics[width=\textwidth]{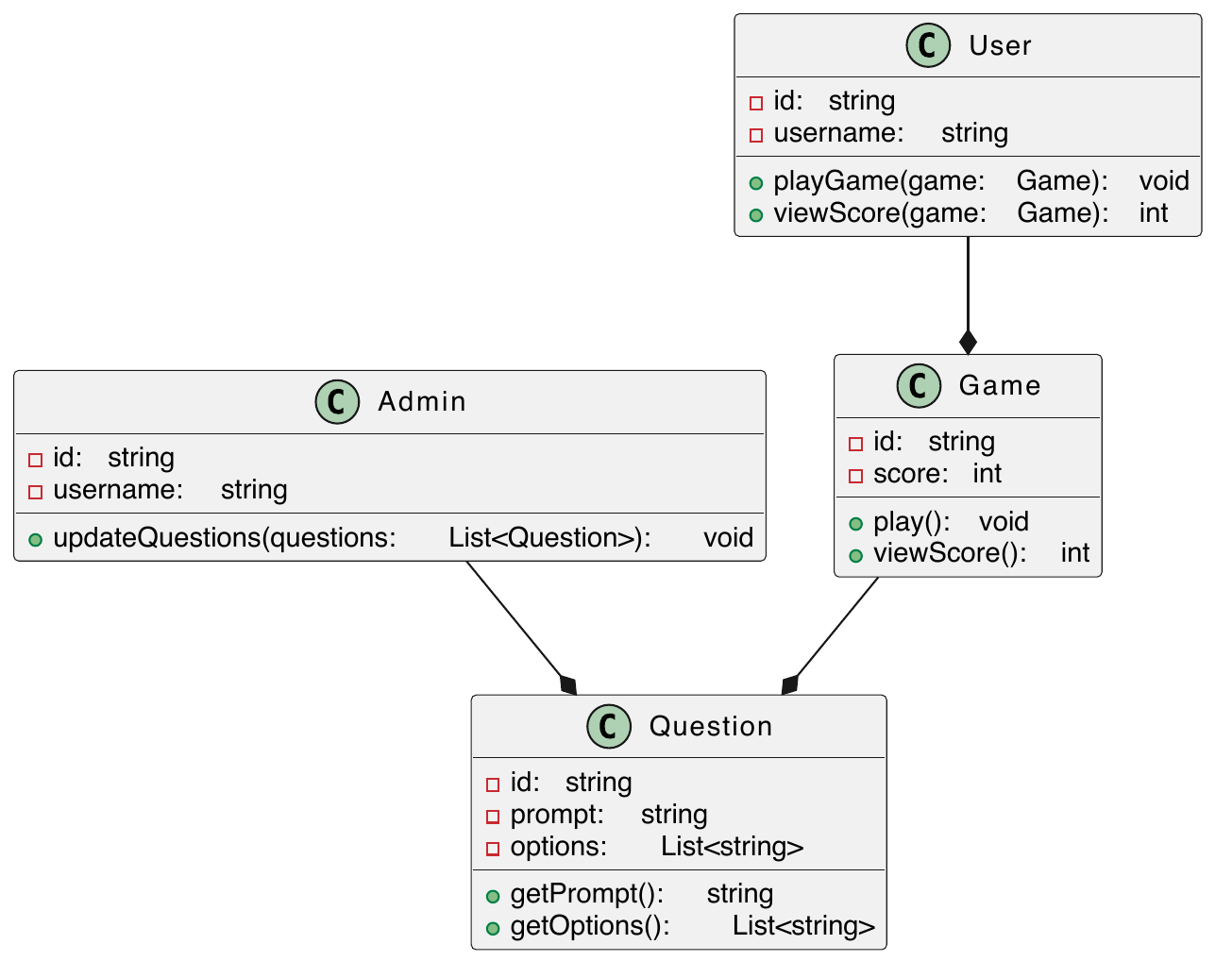}
        \caption{The Class Diagram of SFS Generated by Llama3.3}
        \label{F:Llama_ClassD}
    \end{subfigure}

    \caption{Comparison of class diagrams generated by GPT-5.2, Qwen3.5, DeepSeek-R1, and Llama3.3}
    \label{F:LLM_CD}
\end{figure}

To answer RQ3, we extended our evaluation beyond GPT-5.2 by generating the SFS architecture design with three additional LLMs, i.e., Qwen3.5, DeepSeek-R1 and Llama3.3 (see Section~\ref{sec:LLMSelection}). Likewise, to conduct \textbf{structural and quantitative comparison}, we equipped MAAD with Qwen 3.5(397B), DeepSeek-R1 (671B) and Llama3.3 (70B) as foundational LLMs to generate the architecture using the same SRS input. Figure~\ref{F:Qwen_classD}, Figure~\ref{F:DeepSeek_ClassD} and Figure~\ref{F:Llama_ClassD} depict the class diagrams of SFS generated by Qwen3.5, DeepSeek-R1 and Llama3.3, respectively, which are compared against the GPT-5.2 baseline (Figure~\ref{F:CDmaad}). 

\textbf{Structural Comparison}. The generated class diagrams reveal clear differences in architectural granularity, domain alignment, and subsystem decomposition across the four foundational LLMs. GPT-5.2 produces the most structured and domain-specific architecture, explicitly separating concerns across UI routing (\texttt{ScreenRouter}), game logic (\texttt{StoryEngine}, \texttt{PhysicsEngine}), administrative security (\texttt{AdminAuthService}, \texttt{AuditLogger}), and data persistence (\texttt{QuestionFileRepository}). It also embeds requirements traceability directly into class annotations (e.g., \texttt{ASR-003}, \texttt{FR-021 / NFR-008}), ensuring architectural decisions are grounded in ASRs. Qwen3.5 closely parallels this performance (see Figure~\ref{F:Qwen_classD}); it captures core interactive mechanisms (e.g., \texttt{GameSession} (session management), \texttt{QuestionStore} (question retrieval and storage), \texttt{SceneManager} (scene coordination), \texttt{InputHandler} (user input processing)) and maintains reasonable modularity, yet omits specialized subsystems such as physics simulation and detailed audit logging. 

In contrast, DeepSeek-R1 and Llama3.3 exhibit significant under-specification of architectural structure. DeepSeek-R1 generates a functional but highly abstracted architecture (see Figure~\ref{F:DeepSeek_ClassD}). While it delineates basic operational boundaries (e.g., \texttt{FractionInput}, \texttt{GameState}, \texttt{Question}, \texttt{AuthService}, \texttt{AdminController}), it lacks the intricate subsystem modeling seen in GPT-5.2 and Qwen3.5. Llama3.3 \rev{produces a coarse-grained class model} (see Figure~\ref{F:Llama_ClassD}). It relies on four generic classes (i.e., \texttt{User}, \texttt{Admin}, \texttt{Game}, and \texttt{Question}) with trivial method signatures, entirely missing the domain-specific logic (e.g., fraction arithmetic, physics, and scene rendering) and traceability annotations required for a \rev{project-specific design}. \rev{Table~\ref{tab:llm_distinctive_elements} summarizes these distinctive modeled concerns from the illustrated class diagrams.}

\begin{table*}[htbp]
\centering
\footnotesize
\caption{\rev{Distinctive modeled concerns in the SFS class diagrams across foundational LLMs}}
\label{tab:llm_distinctive_elements}
\setlength{\tabcolsep}{3pt}
{\revcolor
\begin{tabularx}{\textwidth}{@{}>{\RaggedRight\arraybackslash}p{2.6cm} *{4}{>{\RaggedRight\arraybackslash}X}@{}}
\toprule
\textbf{Modeled concern} & \textbf{GPT-5.2} & \textbf{Qwen3.5} & \textbf{DeepSeek-R1} & \textbf{Llama3.3} \\
\midrule
UI / session routing & Present (\texttt{Screen\-Router}) & Present (\texttt{Game\-Session}, \texttt{Scene\-Manager}, \texttt{Input\-Handler}) & Partial (\texttt{Game\-State}) & Coarse (\texttt{User}/\texttt{Game}) \\
Game / domain logic beyond basic Q\&A & Present (\texttt{Story\-Engine}, \texttt{Physics\-Engine}) & Partial (session/scene; no physics) & Partial (\texttt{Fraction\-Input}) & Absent \\
Admin authentication / audit & Present (\texttt{Admin\-Auth\-Service}, \texttt{Audit\-Logger}) & Partial (\texttt{Admin\-User}) & Present (\texttt{Auth\-Service}, \texttt{Admin\-Controller}) & Partial (\texttt{Admin}) \\
Persistence / question store & Present (\texttt{Question\-File\-Repository}) & Present (\texttt{Question\-Store}) & Not explicit & Not explicit \\
Requirement-ID annotations & Present & Present & Limited & Absent \\
\bottomrule
\end{tabularx}}
\end{table*}

\textbf{Quantitative Comparison}. Table~\ref{tab:overall_metrics} presents the architecture-level metrics across the four LLM configurations in MAAD and aligns with the structural differences above. GPT-5.2 and Qwen3.5 exhibit moderate Coupling Degree ($CD$) and Component Coupling Density ($CCD$), which is expected for a multi-component educational system with rich inter-component dependencies. Despite generating more components and interactions, both maintain low Structural Complexity ($SC$: 0.02 and 0.03), indicating sparse dependency graphs rather than densely tangled ones. Conversely, DeepSeek-R1 and Llama3.3 report numerically lower $CD$ and $CCD$; \rev{these lower coupling scores should not be read as superior modularity, because their smaller class sets reduce dependency counts inherently.} Llama3.3's higher Structural Complexity ($SC$=0.17) shows that its few generated elements remain comparatively densely interconnected. GPT-5.2 and Qwen3.5 also have higher Interface Complexity ($IC$: 2.34 and 2.08), reflecting more explicit interface operations in the generated models.

\begin{table}[htbp]
\centering
\caption{Overall Architecture Metrics Comparison across Foundational LLMs for the MAAD framework}
\label{tab:overall_metrics}
\small
\begin{tabular}{lccccccc}
\toprule
\textbf{Model}
& \textbf{CD} & \textbf{Coh} & \textbf{IC} & \textbf{SC} & \textbf{StC} & \textbf{CCD} & \textbf{SMCC} \\
\midrule
GPT-5.2
& 2.13 & 0.50 & 2.34 & 0.02 & 78.80 & 2.53 & 6.20 \\
Qwen3.5
& 1.77 & 0.69 & 2.08 & 0.03 & 65.00 & 2.26 & 5.80 \\
DeepSeek-R1
& 1.56 & 0.67 & 1.56 & 0.03 & 39.30 & 1.99 & 2.20 \\
Llama3.3
& 1.21 & 0.98 & 0.98 & 0.17 & 9.00 & 1.45 & 2.70 \\
\bottomrule
\end{tabular}
\end{table}

Regarding behavioral modeling, State Complexity ($StC$) and State Machine Cyclomatic Complexity ($SMCC$) capture the size and path complexity of the generated state diagrams. GPT-5.2 leads ($StC$=78.80, $SMCC$=6.20), followed by Qwen3.5 ($StC$=65.00, $SMCC$=5.80). DeepSeek-R1 and Llama3.3 yield substantially lower values, consistent with the coarser class models in the structural comparison. Collectively, these metrics show that the choice of foundational LLM changes the structural and behavioral detail of the architectures generated within MAAD: GPT-5.2 and Qwen3.5 produce denser interface and state specifications, whereas DeepSeek-R1 and Llama3.3 produce simpler structures with lower complexity scores.

\begin{tcolorbox}
\rev{\textbf{RQ3 Summary:} \textit{Within the same MAAD workflow, the foundational LLM changes the detail and expressiveness of the generated architecture. GPT-5.2 and Qwen3.5 produce more detailed, requirement-linked designs, whereas DeepSeek-R1 and Llama3.3 produce coarser models. The choice of foundational LLM therefore determines the level of detail in the generated architecture.}}
\end{tcolorbox}

\section{Discussions}\label{sec:Discussions_and_Implications}
In this section, we first explain the experimental results (Section~\ref{sec:Interpretations}) and then outline the implications for subsequent practices and research (Section~\ref{sec:Implications}).
\subsection{Interpretations}\label{sec:Interpretations}
\subsubsection{Interpretations on RQ1 Results}
The results of RQ1 demonstrate that MAAD \rev{automates the generation of structured architectural artifacts---requirements lists, multi-view models, documentation, and evaluation reports---that practitioners treat as useful initial design deliverables.}

\textbf{Specialization and Workflow Alignment Drive Architectural Fidelity.} Unlike general-purpose MAS frameworks (e.g., MetaGPT~\cite{hong2023metagpt}) that simulate broad software company roles and primarily focus on partial artifact generation, MAAD decomposes the design process into specialized cognitive tasks (i.e., analysis, modeling, design, and evaluation) that mirror the workflow of human architects. Moreover, the effectiveness of MAAD stems less from raw generation capability and more from its process-oriented decomposition of architecture design into verifiable stages. This specialization mitigates ``context dilution''~\cite{liu2024lim} by allowing each agent to focus on distinct concerns with tailored prompts and knowledge retrieval strategies.

\textbf{Knowledge Grounding and Iterative Evaluation Mitigate Generative Risks.} A key differentiator between MAAD and the existing MAS frameworks (e.g., MetaGPT~\cite{hong2023metagpt}) is MAAD's integration of external knowledge and the Evaluator agent. The study results highlight that purely generative approaches (e.g., MetaGPT) are prone to hallucinating requirements and neglecting essential structural elements, as evidenced by the zero interface complexity ($IC$=0) across all cases. This result indicates a failure to model component contracts, a cornerstone of architecture design. MAAD addresses this by grounding generation in authoritative knowledge via RAG and enforcing structural rules through iterative evaluation. The integration of the Evaluator agent in the MAAD framework also points to a deeper insight: evaluation is not a post-hoc activity but an integral part of architecture generation. By embedding evaluation into each stage, MAAD effectively transforms architecture design into an iterative process in which evaluation results are used to revise architectural models and documentation. Furthermore, the quantitative metrics (lower structural complexity, higher cohesion) show that MAAD's evaluation mechanism effectively guides the system toward modular and maintainable architecture design, rather than merely producing syntactically valid but architecturally weak outputs.

\textbf{Human-AI Collaboration and the Knowledge Boundary.} The insights from practitioner evaluation underscore that MAAD is best positioned as an augmented intelligent assistant rather than a fully autonomous replacement for human architects. MAAD excels at externalizing explicit knowledge, generating architecture design artifacts (e.g., architectural views, documentation, and evaluation reports), and performing systematic evaluations, thereby reducing manual efforts in artifact construction and consistency checking during architecture design activities. However, the identified challenge in architecture design regarding ``tacit knowledge''~\cite{Tofan2010} reveals a fundamental boundary: architectural design often involves domain-specific nuances, implicit domain assumptions, context-sensitive decisions, and trade-off prioritization that are difficult to capture in specifications or knowledge bases. While MAAD handles the automated synthesis, refinement, and validation of architectural artifacts, human architects remain responsible for architectural strategy, critical decision-making, and balancing competing QAs and business constraints, which often rely on tacit knowledge that current AI cannot fully represent.

\subsubsection{Interpretations on RQ2 Results}
The results of RQ2 demonstrate that external knowledge infusion \rev{changes how MAAD decomposes and represents architectures,} while introducing nuanced trade-offs in metric-based architectural quality.

\textbf{Knowledge Infusion Elevates Abstraction and Mitigates Implementation Bias.} A primary observation from the structural comparison (see Section~\ref{sec:RQ2_Results}) is that \rev{retrieved architectural guidance steers generation away from flat, code-like couplings and toward more explicit role and layer annotations.} Foundation LLMs, predominantly trained on code repositories, tend to default to concrete functional interactions and flat component structures, \rev{as illustrated by the no-RAG configuration's direct controller-to-service couplings.} This result explains the observed trade-off: while abstraction improves modular reasoning and traceability, it can also lead to fragmentation of responsibilities, thereby reducing cohesion. By retrieving authoritative standards and patterns on software architecture design, RAG injects high-level architectural vocabulary (e.g., UML stereotypes, layered abstractions) and design constraints into the reasoning process. This guides the agents to prioritize separation of concerns and contract-based design, \rev{yielding designs with more explicit mediation layers, interface definitions, and requirement annotations in the illustrated case.} The shift from functional dependency mapping to role-based modeling indicates that external knowledge \rev{helps MAAD apply architectural principles rather than only code-level heuristics.} \rev{Retrieved guidance therefore reshapes architectural decomposition and representation; its value depends on matching that guidance to project requirements and QA trade-offs.}

\textbf{Trade-offs Between Decomposition Granularity and Metric-Based Quality.} The quantitative results highlight a critical tension between architectural richness and traditional metric optimization. The results show that RAG promotes finer-grained decomposition, distributing responsibilities across specialized components to enforce modularity. While this fragmentation may lower class-level cohesion metrics, it \rev{can reflect a different allocation of responsibilities rather than a simple loss of relatedness.} Furthermore, the variability in interface and state complexity across projects indicates that knowledge infusion \rev{produces case-dependent structural effects rather than a uniform redesign.} These findings underscore that traditional structural metrics, often optimized for implementation artifacts, may not fully capture the benefits of architectural abstraction, such as improved traceability and standard compliance.

\subsubsection{Interpretations on RQ3 Results}
The comparative evaluation of foundational LLMs within MAAD reveals that MAAD's efficacy is intrinsically bounded by the reasoning capacity and domain fidelity of the underlying LLMs.

\textbf{The Paradox of Complexity Metrics in Generative Design.} A pivotal observation is the potential for standard architectural metrics to be misleading when applied to LLM-generated artifacts. The paradox of complexity metrics in generative design reveals a critical pitfall: low complexity does not equate to high quality when driven by under-specification. Smaller LLMs achieve artificially low $CD$ or $CCD$ scores by omitting domain entities and state transitions, resulting in sparse but semantically incomplete graphs. This finding highlights an important limitation of relying solely on quantitative metrics: without sufficient structural completeness and traceability, such metrics lose their discriminative power and may reward incomplete design.

\textbf{Reasoning Capacity as a Prerequisite for Knowledge Utilization.} The performance disparity across LLMs indicates that the efficacy of knowledge infusion is heavily contingent upon the foundational LLM's reasoning capacity. While all LLMs have access to the same retrieved knowledge, GPT-5.2 and Qwen3.5 demonstrate a superior ability to synthesize this information with the specific constraints of the SRS, producing artifacts that are not only structurally complete but also semantically aligned with domain requirements. In contrast, lower-capability LLMs appear to struggle with integrating retrieved context into coherent, multi-view architecture design, often reverting to generic abstractions. This suggests that in architecture design using MASs, retrieval alone is insufficient; the agent's ability to perform contextual reasoning, cross-view consistency checking and constraint satisfaction over the retrieved knowledge is the limiting factor. MAAD's knowledge mechanisms amplify the capabilities of strong LLMs but cannot fully compensate for the reasoning deficits of weaker ones.

\subsection{Implications}\label{sec:Implications}
Based on the findings of our study, we outline the key implications for the future of AI-driven architecture design.

\rev{\textbf{Connect views to architectural decisions.} Multiple views describe structures and interactions, while decision rationale explains why a proposed design suits its context~\cite{Kruchten2009DecisionView}. MAAD provides explicit requirements, model, and documentation artifacts through which these links can be inspected. In practice, an architectural decision should identify the relevant requirements, alternatives, assumptions, expected consequences, and the reason for selecting an option. This makes traceability useful for evaluating a decision, beyond locating a requirement identifier.}

\textbf{Integrated Evaluation Mechanisms and Traceability Matter.} The results indicate that the main limitation of existing LLM-based or MAS approaches may not be insufficient model capability, but rather the absence of structured intermediate representations of architectural constraints and designs, together with control mechanisms for regulating the generative process. End-to-end generative pipelines inherently suffer from uncertainty propagation, which can be effectively mitigated by enforcing explicit requirements categorization and iterative validation. Incorporating dedicated evaluation agents with iterative feedback loops is critical for ensuring architectural quality. Research should advance evaluation criteria, including dynamic QA analysis and automated conformance checking. Moreover, MAAD demonstrates that embedding traceability links directly into architectural models enhances the transparency and validation of LLM-generated design decisions against stakeholder requirements. Therefore, it is worthwhile for researchers to further explore standardized mechanisms for maintaining end-to-end traceability across AI-generated artifacts to better support architecture design and validation processes.

\textbf{Architectural Quality is Multi-Dimensional and Requires Controlled Knowledge Integration.} Our findings indicate that architectural quality extends beyond structural or syntactic properties, yet conventional metrics fail to fully capture the value of knowledge-driven design. To address this limitation, researchers and practitioners can explore more comprehensive evaluation frameworks that explicitly incorporate dimensions such as traceability, explainability, and standards compliance, which are essential for assessing how external knowledge translates into architectural integrity. However, knowledge integration must be carefully regulated: unconstrained injection of external information can inadvertently introduce over-fragmentation, unnecessary complexity, or misaligned abstractions. Consequently, adaptive mechanisms are required to dynamically govern the timing, scope, and granularity of knowledge application during the architecture design process. Moreover, the inherent trade-offs introduced by RAG-based approaches necessitate human-in-the-loop oversight. Experienced architects remain indispensable for validating abstraction levels, mitigating over-engineering risks, and ensuring that architecture design remains tightly aligned with project-specific constraints and requirements. Together, these insights underscore the need for holistic evaluation paradigms and a balanced, human-guided methodology for knowledge-augmented architecture design.

\textbf{LLM Selection in MAAD Critically Shapes Architectural Outcomes.} Our findings reveal that multi-agent collaboration cannot fully compensate for the intrinsic limitations of the underlying foundation LLMs. Although mechanisms such as role specialization and iterative feedback improve structural coherence and enable incremental architecture refinement, they remain bounded by the semantic understanding and reasoning capacity of the base models. As a result, MAAD cannot recover foundational architectural principles or domain-specific knowledge that were not captured during initial generation. Rather than functioning as a corrective mechanism, multi-agent orchestration acts as a capability amplifier: it elevates the generation precision and reasoning depth of stronger LLMs, while exposing the deficiencies of weaker LLMs more apparently. This demonstrates that the base LLM's competence might establish a performance ceiling for the entire MAAD pipeline, rather than a mere bottleneck that coordination strategies can easily bypass. For researchers, this underscores the necessity of capability-aware orchestration, including dynamic model routing, LLM-specific interaction protocols, and hybrid selection strategies that align agent roles with model strengths. For practitioners, it highlights that selecting appropriate base LLMs is a prerequisite to designing an effective MAS for architecture design. Without this alignment, increasing multi-agent orchestration is unlikely to overcome foundational design flaws and may instead result in fragmented or shallow architectural designs.

\textbf{Human Validation Remains Essential for Architecture Decisions.} Expert interviews consistently positioned MAAD as ``undoubtedly useful for assisting architects, especially mismatch reports'' while emphasizing that explainability issues persist. Practitioners appreciated MAAD's ability to comprehensively recall architectural design knowledge and systematically analyze the architectural design process, yet emphasized the continued need for human validation of the generated artifacts, particularly noting that ``architects often favor their own solutions over those generated by LLMs''. This feedback demonstrates that practitioners should position automated architecture design tools as assistant technologies requiring human validation rather than autonomous systems replacing architects, especially in safety-critical domains where architectural decisions have significant consequences.

\rev{\textbf{Integrate autonomous execution with human oversight.} MAAD automates artifact generation and iterative agent feedback; human involvement concerns the validation and adoption of the resulting design. Consistent with Ivers and Ozkaya~\cite{Ivers2025AutomationGap}, we recommend four review points for integrating MAAD into an architecting workflow. First, stakeholders confirm requirements interpretations, ASRs, priorities, and unresolved ambiguities. Second, the architect reviews candidate decompositions and quality-attribute scenarios against business and technical constraints. Third, domain experts inspect the assembled documentation, traceability, and reported risks, using prototypes or specialized analyses where necessary. Fourth, the responsible architect and stakeholders approve the selected decisions and residual risks.}

\rev{\textbf{Three memory layers support iterative context, design history, and architectural knowledge reuse.} Working memory keeps the current revision context, episodic memory records each stage's design history, and semantic memory stores reusable architectural guidance that later tasks can retrieve (Section~\ref{sec:MemoryMechanism}). These roles of the three memory layers explain how evaluation feedback and prior architectural decisions move through the MAAD workflow.}

\section{Threats to Validity}\label{sec:Threats}
In this section, we discuss the potential threats to the validity of our study and the measures adopted to mitigate them, following the guidelines of Wohlin \textit{et al.}~\cite{Wohlin2012ESE}.

\rev{\textbf{Internal validity} concerns whether observed differences between configurations can be attributed to the intended factors. In RQ2, the intended difference is the presence (with-RAG) or absence (no-RAG) of retrieved external knowledge under shared base prompts and agent workflow; in RQ3, it is the foundational LLM under the same workflow. Residual factors include model versions, retrieved external-knowledge contents, and possible memory carry-over across cases. To mitigate these threats, we keep the workflow fixed when ablating external-knowledge retrieval, set decoding temperature to $0.0$ for all compared LLMs, and interpret differences at the configuration level under these controls.}

\textbf{Construct validity} in this study concerns whether MAAD can generate architectural artifacts that correctly reflect requirements, design rationale, and architectural constraints. Potential threats arise from inherent LLM limitations, such as hallucinations~\cite{Zhang2023ssai} and inconsistent reasoning, which may lead to incorrect architectural decisions or inconsistencies among generated artifacts. To mitigate these threats, MAAD employs a multi-agent cross-verification workflow, strict JSON/UML output constraints, and an integrated Evaluator agent for rule-based consistency checking. Through these mechanisms, the correctness and traceability of the generated architecture artifacts can be further improved. \rev{In addition, the structural metrics used in the evaluation describe selected properties of the generated architectural representations rather than overall architectural fitness for a project context; we therefore interpret these metrics alongside artifact inspection and practitioner feedback.}

\textbf{External validity} concerns the generalizability of MAAD across software architecture design scenarios. Potential threats might stem from the diversity of SRSs, the coverage of architectural knowledge, and the selected LLMs. Although our evaluation includes ten SRSs from different domains and grounds MAAD in established architecture standards (e.g., ISO/IEC/IEEE 42010), some domain-specific architectural practices~\cite{hofmeister2007general} may still be underrepresented. Moreover, the study results may vary when different LLMs are used. To alleviate these threats, we intentionally selected projects from multiple domains and adopted authoritative architecture knowledge sources to improve the representativeness of the evaluation.

\textbf{Conclusion validity} pertains to whether the evaluation results accurately reflect the effectiveness of MAAD in supporting software architecture design. A major threat lies in the subjectivity of architectural quality assessment, as no single metric can fully capture architectural quality. To mitigate this threat, we complemented quantitative metrics with qualitative evaluations, including ATAM-based evaluation reports, mismatch analysis, and semi-structured interviews with experienced practitioners. This combination of architectural quality evaluation methods reduces reliance on any single evaluation method and increases confidence in the findings.

\section{Conclusions and Future Work}\label{sec:Conclusion}
In this paper, we propose MAAD, a knowledge-driven multi-agent framework that automates software architecture design through collaborative requirement analysis, multi-view architecture modeling, architecture documentation synthesis, and systematic evaluation. MAAD comprises four specialized agents (i.e., Analyst, Modeler, Designer, and Evaluator agent) that collaboratively generate software architecture designs from given software requirements specifications.

MAAD demonstrates that knowledge-driven multi-agent orchestration can elevate the automation of architectural artifact generation. Our findings also show the limits of that automation: the framework's efficacy is constrained by the base LLM's reasoning capacity, and knowledge infusion requires calibration to project context. \rev{Our empirical evaluation shows that, compared with the selected MetaGPT outputs, MAAD produces more detailed architectural descriptions and preliminary evaluation reports, with differences in the reported structural indicators.} Furthermore, the infusion of external knowledge changes design abstraction and representation, though its impact on traditional structural metrics is context-dependent. Crucially, the reasoning capacity of the underlying LLM bounds the framework's performance, with high-capability models (GPT-5.2, Qwen3.5) producing more detailed, domain-aligned architectures, whereas smaller models yield coarser designs. \rev{Practitioner interviews identify the generated models, documents, and evaluation reports as useful initial design deliverables. These findings support autonomous artifact generation within an architecting process in which stakeholders validate requirements and architects review and approve design decisions.}

Based on these insights, we envision several directions for future work. First, researchers can develop adaptive knowledge retrieval and routing mechanisms that dynamically calibrate the granularity and scope of retrieved architectural patterns to prevent over-fragmentation of architectural knowledge and preserve design coherence. \rev{Second, it is promising to explore hybrid LLM orchestration strategies (e.g., combining reasoning-focused and lightweight models like Jev/Kev/Laya) to align agent roles with model strengths}, enabling fast and robust cross-domain generalization of architecture generation capabilities. Third, dynamic QA simulation tools (e.g., performance profiling, security scanning) can be integrated into the Evaluator agent to provide real-time, quantifiable feedback during architecture generation. \rev{Fourth, future evaluation should measure the cost of architecture artifact generation and human review, including retrieval context size, token consumption, and human review effort.} Finally, recognizing the enduring role of tacit knowledge and contextual judgment in software architecture design and decision-making, we aim to keep refining MAAD to support interactive co-design, constructing quick-validation coding agents to evaluate architecture design in the implementation phase, ensuring that AI-assisted architecture remains transparent, explainable, and aligned with project-specific constraints in safety-critical and enterprise environments.

% Make % a normal character so href targets embed a real "%20" (not TeX comment / \040).
\begingroup
\catcode`\%=12
\gdef\MAADSFSAnalystURL{https://github.com/RuiyinL/MAAD/tree/main/MAAD%20artifacts/RQ1_Results/MAAD_Output/SFS/Analyst}
\gdef\MAADSFSDesignerURL{https://github.com/RuiyinL/MAAD/tree/main/MAAD%20artifacts/RQ1_Results/MAAD_Output/SFS/Designer}
\gdef\MAADSFSEvaluatorURL{https://github.com/RuiyinL/MAAD/tree/main/MAAD%20artifacts/RQ1_Results/MAAD_Output/SFS/Evaluator}
\endgroup

\section*{Data Availability}\label{sec:DataAvailability}
\rev{The replication package is available at~\cite{onlinepackage_TOSEM}. It provides the requirements dataset, complete agent prompts, and generated artifacts organized by research question, configuration, case, and agent. The SFS example can be inspected through the \href{\MAADSFSAnalystURL}{\textit{Analyst} requirements}, \href{\MAADSFSDesignerURL}{\textit{Designer} documentation}, and \href{\MAADSFSEvaluatorURL}{\textit{Evaluator} reports}. These links identify the publicly available outputs; implementation and execution details should be replicated together with the corresponding experiment configuration.}

\section*{Acknowledgments}\label{sec:Acknowledgments}
This work has been partially supported by the National Natural Science Foundation of China (NSFC) with Grant Nos. 92582203 and 62402348.
%, and the Major Science and Technology Project of Hubei Province with Grant No. 2024BAA008. The numerical calculations in this paper have been done on the supercomputing system in the Supercomputing Center of Wuhan University.

% \begin{acks}
% This research is supported by the National Natural Science Foundation of China (NSFC) with Grant No. 62402348 and 62172311; National Research Foundation, Prime Minister's Office, Singapore under the Campus for Research Excellence and Technological Enterprise (CREATE) programme; the National Research Foundation, Singapore, and DSO National Laboratories under the AI Singapore Programme (AISG Award No: AISG2-GC-2023-008). Besides, the authors would like to thank all participants in this study.
% \end{acks}

% \clearpage
\bibliographystyle{ACM-Reference-Format}
\bibliography{TOSEM_ref}

%\end{sloppypar}
\end{document}